\documentclass[12pt]{article}
\usepackage{graphicx}   
\usepackage{xcolor}      

\usepackage[english]{babel} 
\usepackage[T1]{fontenc}    
\usepackage[ansinew]{inputenc} 
\usepackage{lmodern} 
\usepackage{microtype} 
\usepackage[condensed]{iwona}

\usepackage{amsmath} 
\usepackage{amssymb} 
\usepackage{algorithm}  
\usepackage{algpseudocode}

\usepackage[switch,running,right,mathlines]{lineno}  
\usepackage{csquotes} 

\usepackage[matstyle=bbold]{MattsMacros}

\usepackage{titling}  
\usepackage{url}  
\usepackage[style=authoryear,backend=biber,firstinits=false,maxcitenames=2,maxbibnames=99,urldate=iso8601,uniquename=false,uniquelist=false,url=false]{biblatex}

\bibliography{examples}
\makeatletter
\def\blx@bblfile@biber{%
  \blx@secinit
  \begingroup
  \blx@bblstart
\begingroup
\makeatletter
\@ifundefined{ver@biblatex.sty}
  {\@latex@error
     {Missing 'biblatex' package}
     {The bibliography requires the 'biblatex' package.}
      \aftergroup }
  {}
\endgroup

\refsection{0}
  \sortlist{entry}{nyt}
    \entry{ait2008cross}{article}{}
      \name{labelname}{4}{}{%
        {{hash=8315fcd56c6c4b8578e8fcd68d45f264}{{Ait-Sa{\"i}di}}{A\bibinitperiod}{A.}{A\bibinitperiod}{}{}{}{}}%
        {{hash=e19d2a92dc7a4091d092b2e8b12b52ba}{Ferraty}{F\bibinitperiod}{F.}{F\bibinitperiod}{}{}{}{}}%
        {{hash=bf8b6c1e2f7e0e3edde6be5b155fd4fd}{Kassa}{K\bibinitperiod}{R.}{R\bibinitperiod}{}{}{}{}}%
        {{hash=919b238b28abef467c8d352e1bd54579}{Vieu}{V\bibinitperiod}{P.}{P\bibinitperiod}{}{}{}{}}%
      }
      \name{author}{4}{}{%
        {{hash=8315fcd56c6c4b8578e8fcd68d45f264}{{Ait-Sa{\"i}di}}{A\bibinitperiod}{A.}{A\bibinitperiod}{}{}{}{}}%
        {{hash=e19d2a92dc7a4091d092b2e8b12b52ba}{Ferraty}{F\bibinitperiod}{F.}{F\bibinitperiod}{}{}{}{}}%
        {{hash=bf8b6c1e2f7e0e3edde6be5b155fd4fd}{Kassa}{K\bibinitperiod}{R.}{R\bibinitperiod}{}{}{}{}}%
        {{hash=919b238b28abef467c8d352e1bd54579}{Vieu}{V\bibinitperiod}{P.}{P\bibinitperiod}{}{}{}{}}%
      }
      \list{publisher}{1}{%
        {Taylor \& Francis}%
      }
      \strng{namehash}{8c72144456573401a1b720b6de872e64}
      \strng{fullhash}{1b111fd8bd787827f7780fbdafed66ca}
      \field{sortinit}{A}
      \field{labelyear}{2008}
      \field{labeltitle}{Cross-validated estimations in the single-functional index model}
      \field{journaltitle}{Statistics}
      \field{number}{6}
      \field{title}{Cross-validated estimations in the single-functional index model}
      \field{volume}{42}
      \field{year}{2008}
      \field{pages}{475\bibrangedash 494}
    \endentry
    \entry{hooker2012functional}{article}{}
      \name{labelname}{3}{}{%
        {{hash=6d5d57643c8cfd202ae78d1607c54cf6}{Asencio}{A\bibinitperiod}{M.}{M\bibinitperiod}{}{}{}{}}%
        {{hash=b4534ebe5315666e28756f054a7a003c}{Hooker}{H\bibinitperiod}{G.}{G\bibinitperiod}{}{}{}{}}%
        {{hash=d904c9c46ed9a8c435a23188e3fb631c}{Gao}{G\bibinitperiod}{H.\bibnamedelimi O.}{H\bibinitperiod\bibinitdelim O\bibinitperiod}{}{}{}{}}%
      }
      \name{author}{3}{}{%
        {{hash=6d5d57643c8cfd202ae78d1607c54cf6}{Asencio}{A\bibinitperiod}{M.}{M\bibinitperiod}{}{}{}{}}%
        {{hash=b4534ebe5315666e28756f054a7a003c}{Hooker}{H\bibinitperiod}{G.}{G\bibinitperiod}{}{}{}{}}%
        {{hash=d904c9c46ed9a8c435a23188e3fb631c}{Gao}{G\bibinitperiod}{H.\bibnamedelimi O.}{H\bibinitperiod\bibinitdelim O\bibinitperiod}{}{}{}{}}%
      }
      \strng{namehash}{69726da54241606e3f03cdb3e331825c}
      \strng{fullhash}{bdfae83fe6bed18286229c355aa98985}
      \field{sortinit}{A}
      \field{labelyear}{2013}
      \field{labeltitle}{Functional Convolution Models}
      \field{journaltitle}{Statistical Modelling}
      \field{pubstate}{In press}
      \field{title}{Functional Convolution Models}
      \field{urlday}{06}
      \field{urlmonth}{10}
      \field{urlyear}{2013}
      \field{year}{2013}
      \verb{url}
      \verb http://www.bscb.cornell.edu/~hooker/EmissionsPaper.pdf
      \endverb
    \endentry
    \entry{bates2013lme4}{manual}{}
      \name{labelname}{3}{}{%
        {{hash=4a4417f954b766f602defea7a0feda7b}{Bates}{B\bibinitperiod}{D.}{D\bibinitperiod}{}{}{}{}}%
        {{hash=f1ff1a9390e6820033fc7a228221e414}{Maechler}{M\bibinitperiod}{M.}{M\bibinitperiod}{}{}{}{}}%
        {{hash=fc0ee7abee741b5fbd85cdc1d6ff410e}{Bolker}{B\bibinitperiod}{B.}{B\bibinitperiod}{}{}{}{}}%
      }
      \name{author}{3}{}{%
        {{hash=4a4417f954b766f602defea7a0feda7b}{Bates}{B\bibinitperiod}{D.}{D\bibinitperiod}{}{}{}{}}%
        {{hash=f1ff1a9390e6820033fc7a228221e414}{Maechler}{M\bibinitperiod}{M.}{M\bibinitperiod}{}{}{}{}}%
        {{hash=fc0ee7abee741b5fbd85cdc1d6ff410e}{Bolker}{B\bibinitperiod}{B.}{B\bibinitperiod}{}{}{}{}}%
      }
      \strng{namehash}{a0837773990253a159a0950c429e8842}
      \strng{fullhash}{ec6e186f2c35c1d5c3e3b83a9b0e61ea}
      \field{sortinit}{B}
      \field{labelyear}{2013}
      \field{labeltitle}{{lme}4: Linear mixed-effects models using {S}4 classes, {R} package version 1.0-4}
      \field{title}{{lme}4: Linear mixed-effects models using {S}4 classes, {R} package version 1.0-4}
      \field{urlday}{06}
      \field{urlmonth}{10}
      \field{urlyear}{2013}
      \field{year}{2013}
      \verb{url}
      \verb http://CRAN.R-project.org/package=lme4
      \endverb
    \endentry
    \entry{cardot2003testing}{article}{}
      \name{labelname}{4}{}{%
        {{hash=9dfa31c3742cff6c6149d2feeaf3670b}{Cardot}{C\bibinitperiod}{H.}{H\bibinitperiod}{}{}{}{}}%
        {{hash=e19d2a92dc7a4091d092b2e8b12b52ba}{Ferraty}{F\bibinitperiod}{F.}{F\bibinitperiod}{}{}{}{}}%
        {{hash=be0190a2091c652d8de35ca3953706cd}{Mas}{M\bibinitperiod}{A.}{A\bibinitperiod}{}{}{}{}}%
        {{hash=205fc242ba6047061b4bcc83780d20cd}{Sarda}{S\bibinitperiod}{P.}{P\bibinitperiod}{}{}{}{}}%
      }
      \name{author}{4}{}{%
        {{hash=9dfa31c3742cff6c6149d2feeaf3670b}{Cardot}{C\bibinitperiod}{H.}{H\bibinitperiod}{}{}{}{}}%
        {{hash=e19d2a92dc7a4091d092b2e8b12b52ba}{Ferraty}{F\bibinitperiod}{F.}{F\bibinitperiod}{}{}{}{}}%
        {{hash=be0190a2091c652d8de35ca3953706cd}{Mas}{M\bibinitperiod}{A.}{A\bibinitperiod}{}{}{}{}}%
        {{hash=205fc242ba6047061b4bcc83780d20cd}{Sarda}{S\bibinitperiod}{P.}{P\bibinitperiod}{}{}{}{}}%
      }
      \list{publisher}{1}{%
        {Wiley Online Library}%
      }
      \strng{namehash}{bec91794ae5a7c369d87a8acb0ce7c61}
      \strng{fullhash}{224c0044bd26b0ea8ce78d463a143ad6}
      \field{sortinit}{C}
      \field{labelyear}{2003}
      \field{labeltitle}{Testing hypotheses in the functional linear model}
      \field{journaltitle}{Scandinavian Journal of Statistics}
      \field{number}{1}
      \field{title}{Testing hypotheses in the functional linear model}
      \field{volume}{30}
      \field{year}{2003}
      \field{pages}{241\bibrangedash 255}
    \endentry
    \entry{chen2011single}{article}{}
      \name{labelname}{3}{}{%
        {{hash=20cca183f63a6c0e940c18ba4a81deac}{Chen}{C\bibinitperiod}{D.}{D\bibinitperiod}{}{}{}{}}%
        {{hash=f16a12039f3f5853fa29622054cce443}{Hall}{H\bibinitperiod}{P.}{P\bibinitperiod}{}{}{}{}}%
        {{hash=26ae59136e82514e73fdff084edf19c7}{M{\"u}ller}{M\bibinitperiod}{H.\bibnamedelimi G.}{H\bibinitperiod\bibinitdelim G\bibinitperiod}{}{}{}{}}%
      }
      \name{author}{3}{}{%
        {{hash=20cca183f63a6c0e940c18ba4a81deac}{Chen}{C\bibinitperiod}{D.}{D\bibinitperiod}{}{}{}{}}%
        {{hash=f16a12039f3f5853fa29622054cce443}{Hall}{H\bibinitperiod}{P.}{P\bibinitperiod}{}{}{}{}}%
        {{hash=26ae59136e82514e73fdff084edf19c7}{M{\"u}ller}{M\bibinitperiod}{H.\bibnamedelimi G.}{H\bibinitperiod\bibinitdelim G\bibinitperiod}{}{}{}{}}%
      }
      \list{publisher}{1}{%
        {Institute of Mathematical Statistics}%
      }
      \strng{namehash}{55d035f29522a9b082db07c6065e7d7c}
      \strng{fullhash}{b1000d55593a731940f9b28223b3c919}
      \field{sortinit}{C}
      \field{labelyear}{2011}
      \field{labeltitle}{Single and multiple index functional regression models with nonparametric link}
      \field{journaltitle}{The Annals of Statistics}
      \field{number}{3}
      \field{title}{Single and multiple index functional regression models with nonparametric link}
      \field{volume}{39}
      \field{year}{2011}
      \field{pages}{1720\bibrangedash 1747}
    \endentry
    \entry{clark2007emissions}{report}{}
      \name{labelname}{5}{}{%
        {{hash=3d1926990c40d9f729295479f32b1e1e}{Clark}{C\bibinitperiod}{N.\bibnamedelimi N.}{N\bibinitperiod\bibinitdelim N\bibinitperiod}{}{}{}{}}%
        {{hash=9d2a52250d2a3eeab7cebcdd76aba720}{Gautam}{G\bibinitperiod}{M.}{M\bibinitperiod}{}{}{}{}}%
        {{hash=6e9205ec9ccd540e1dde590eebcac869}{Wayne}{W\bibinitperiod}{W.\bibnamedelimi S.}{W\bibinitperiod\bibinitdelim S\bibinitperiod}{}{}{}{}}%
        {{hash=47ba1ff3bda0bb3546f893d7bfb8d2d2}{Lyons}{L\bibinitperiod}{G.\bibnamedelimi W.}{G\bibinitperiod\bibinitdelim W\bibinitperiod}{}{}{}{}}%
        {{hash=63e6264f5a5e6bf837fe9d06501700a9}{Thompson}{T\bibinitperiod}{G.\bibnamedelimi J.}{G\bibinitperiod\bibinitdelim J\bibinitperiod}{}{}{}{}}%
      }
      \name{author}{5}{}{%
        {{hash=3d1926990c40d9f729295479f32b1e1e}{Clark}{C\bibinitperiod}{N.\bibnamedelimi N.}{N\bibinitperiod\bibinitdelim N\bibinitperiod}{}{}{}{}}%
        {{hash=9d2a52250d2a3eeab7cebcdd76aba720}{Gautam}{G\bibinitperiod}{M.}{M\bibinitperiod}{}{}{}{}}%
        {{hash=6e9205ec9ccd540e1dde590eebcac869}{Wayne}{W\bibinitperiod}{W.\bibnamedelimi S.}{W\bibinitperiod\bibinitdelim S\bibinitperiod}{}{}{}{}}%
        {{hash=47ba1ff3bda0bb3546f893d7bfb8d2d2}{Lyons}{L\bibinitperiod}{G.\bibnamedelimi W.}{G\bibinitperiod\bibinitdelim W\bibinitperiod}{}{}{}{}}%
        {{hash=63e6264f5a5e6bf837fe9d06501700a9}{Thompson}{T\bibinitperiod}{G.\bibnamedelimi J.}{G\bibinitperiod\bibinitdelim J\bibinitperiod}{}{}{}{}}%
      }
      \list{institution}{1}{%
        {Coordinating Research Council, Inc. (CRC)}%
      }
      \list{location}{1}{%
        {Alpharetta, GA}%
      }
      \strng{namehash}{d2e3e05ccc92472130a821eb2e901ad8}
      \strng{fullhash}{7ff44ca1b5cdc181dc4744f8f0607266}
      \field{sortinit}{C}
      \field{labelyear}{2007}
      \field{labeltitle}{Heavy-duty vehicle chassis dynamometer testing for emissions inventory, air quality modeling, source apportionment, and air toxics emissions inventory}
      \field{number}{{CRC Rep. No. E55/59}}
      \field{title}{Heavy-duty vehicle chassis dynamometer testing for emissions inventory, air quality modeling, source apportionment, and air toxics emissions inventory}
      \field{type}{techreport}
      \field{urlday}{06}
      \field{urlmonth}{10}
      \field{urlyear}{2013}
      \field{year}{2007}
      \verb{url}
      \verb http://www.crcao.com/reports/recentstudies2007/E-55-59/E-55_59_Final_Report_23AUG2007.pdf
      \endverb
    \endentry
    \entry{clark2010expressing}{article}{}
      \name{labelname}{6}{}{%
        {{hash=3d1926990c40d9f729295479f32b1e1e}{Clark}{C\bibinitperiod}{N.\bibnamedelimi N.}{N\bibinitperiod\bibinitdelim N\bibinitperiod}{}{}{}{}}%
        {{hash=7fa901a692bf6b65ea039cb97ae096f9}{Vora}{V\bibinitperiod}{K.\bibnamedelimi A.}{K\bibinitperiod\bibinitdelim A\bibinitperiod}{}{}{}{}}%
        {{hash=2126ff11704e4078717686170cbb1b15}{Wang}{W\bibinitperiod}{L.}{L\bibinitperiod}{}{}{}{}}%
        {{hash=9d2a52250d2a3eeab7cebcdd76aba720}{Gautam}{G\bibinitperiod}{M.}{M\bibinitperiod}{}{}{}{}}%
        {{hash=6e9205ec9ccd540e1dde590eebcac869}{Wayne}{W\bibinitperiod}{W.\bibnamedelimi S.}{W\bibinitperiod\bibinitdelim S\bibinitperiod}{}{}{}{}}%
        {{hash=63e6264f5a5e6bf837fe9d06501700a9}{Thompson}{T\bibinitperiod}{G.\bibnamedelimi J.}{G\bibinitperiod\bibinitdelim J\bibinitperiod}{}{}{}{}}%
      }
      \name{author}{6}{}{%
        {{hash=3d1926990c40d9f729295479f32b1e1e}{Clark}{C\bibinitperiod}{N.\bibnamedelimi N.}{N\bibinitperiod\bibinitdelim N\bibinitperiod}{}{}{}{}}%
        {{hash=7fa901a692bf6b65ea039cb97ae096f9}{Vora}{V\bibinitperiod}{K.\bibnamedelimi A.}{K\bibinitperiod\bibinitdelim A\bibinitperiod}{}{}{}{}}%
        {{hash=2126ff11704e4078717686170cbb1b15}{Wang}{W\bibinitperiod}{L.}{L\bibinitperiod}{}{}{}{}}%
        {{hash=9d2a52250d2a3eeab7cebcdd76aba720}{Gautam}{G\bibinitperiod}{M.}{M\bibinitperiod}{}{}{}{}}%
        {{hash=6e9205ec9ccd540e1dde590eebcac869}{Wayne}{W\bibinitperiod}{W.\bibnamedelimi S.}{W\bibinitperiod\bibinitdelim S\bibinitperiod}{}{}{}{}}%
        {{hash=63e6264f5a5e6bf837fe9d06501700a9}{Thompson}{T\bibinitperiod}{G.\bibnamedelimi J.}{G\bibinitperiod\bibinitdelim J\bibinitperiod}{}{}{}{}}%
      }
      \list{publisher}{1}{%
        {ACS Publications}%
      }
      \strng{namehash}{d2e3e05ccc92472130a821eb2e901ad8}
      \strng{fullhash}{22fdd6f113a7a9946de66a93c3cf3f0f}
      \field{sortinit}{C}
      \field{labelyear}{2010}
      \field{labeltitle}{Expressing cycles and their emissions on the basis of properties and results from other cycles}
      \field{journaltitle}{Environmental science \& technology}
      \field{number}{15}
      \field{title}{Expressing cycles and their emissions on the basis of properties and results from other cycles}
      \field{volume}{44}
      \field{year}{2010}
      \field{pages}{5986\bibrangedash 5992}
    \endentry
    \entry{refund}{manual}{}
      \name{labelname}{6}{}{%
        {{hash=6dac3bc879f150e7889487a5f8816d26}{Crainiceanu}{C\bibinitperiod}{C.\bibnamedelimi M.}{C\bibinitperiod\bibinitdelim M\bibinitperiod}{}{}{}{}}%
        {{hash=d888d5b7a3ceef7e6f4be790a8d2a000}{Reiss}{R\bibinitperiod}{P.\bibnamedelimi T.}{P\bibinitperiod\bibinitdelim T\bibinitperiod}{}{}{}{}}%
        {{hash=947b47a6cf08079790bb15f78caa2124}{Goldsmith}{G\bibinitperiod}{J.}{J\bibinitperiod}{}{}{}{}}%
        {{hash=3c7826b026033c98b5b25a634e1b63fa}{Huang}{H\bibinitperiod}{L.}{L\bibinitperiod}{}{}{}{}}%
        {{hash=69a67147454fa7cff7d39a586d509695}{Huo}{H\bibinitperiod}{L.}{L\bibinitperiod}{}{}{}{}}%
        {{hash=4f1e5e96ac9e3d1124e4f0be96e8c9f6}{Scheipl}{S\bibinitperiod}{F.}{F\bibinitperiod}{}{}{}{}}%
      }
      \name{author}{6}{}{%
        {{hash=6dac3bc879f150e7889487a5f8816d26}{Crainiceanu}{C\bibinitperiod}{C.\bibnamedelimi M.}{C\bibinitperiod\bibinitdelim M\bibinitperiod}{}{}{}{}}%
        {{hash=d888d5b7a3ceef7e6f4be790a8d2a000}{Reiss}{R\bibinitperiod}{P.\bibnamedelimi T.}{P\bibinitperiod\bibinitdelim T\bibinitperiod}{}{}{}{}}%
        {{hash=947b47a6cf08079790bb15f78caa2124}{Goldsmith}{G\bibinitperiod}{J.}{J\bibinitperiod}{}{}{}{}}%
        {{hash=3c7826b026033c98b5b25a634e1b63fa}{Huang}{H\bibinitperiod}{L.}{L\bibinitperiod}{}{}{}{}}%
        {{hash=69a67147454fa7cff7d39a586d509695}{Huo}{H\bibinitperiod}{L.}{L\bibinitperiod}{}{}{}{}}%
        {{hash=4f1e5e96ac9e3d1124e4f0be96e8c9f6}{Scheipl}{S\bibinitperiod}{F.}{F\bibinitperiod}{}{}{}{}}%
      }
      \strng{namehash}{dbc5c1dd5321fcadd62a7e6d440c82ea}
      \strng{fullhash}{292e057ee1f5f261b2411a446bfce749}
      \field{sortinit}{C}
      \field{labelyear}{2013}
      \field{labeltitle}{{refund}: Regression with Functional Data, {R} package version 0.1-7}
      \field{title}{{refund}: Regression with Functional Data, {R} package version 0.1-7}
      \field{urlday}{06}
      \field{urlmonth}{10}
      \field{urlyear}{2013}
      \field{year}{2013}
      \verb{url}
      \verb http://CRAN.R-project.org/package=refund
      \endverb
    \endentry
    \entry{crainiceanu2004Likelihood}{article}{}
      \name{labelname}{2}{}{%
        {{hash=6dac3bc879f150e7889487a5f8816d26}{Crainiceanu}{C\bibinitperiod}{C.\bibnamedelimi M.}{C\bibinitperiod\bibinitdelim M\bibinitperiod}{}{}{}{}}%
        {{hash=6c4cb8ccb3c3853c45b3e9179df4dea2}{Ruppert}{R\bibinitperiod}{D.}{D\bibinitperiod}{}{}{}{}}%
      }
      \name{author}{2}{}{%
        {{hash=6dac3bc879f150e7889487a5f8816d26}{Crainiceanu}{C\bibinitperiod}{C.\bibnamedelimi M.}{C\bibinitperiod\bibinitdelim M\bibinitperiod}{}{}{}{}}%
        {{hash=6c4cb8ccb3c3853c45b3e9179df4dea2}{Ruppert}{R\bibinitperiod}{D.}{D\bibinitperiod}{}{}{}{}}%
      }
      \list{publisher}{1}{%
        {Wiley Online Library}%
      }
      \strng{namehash}{71e2c7ec4b975c9bc608d8b62b5987c2}
      \strng{fullhash}{71e2c7ec4b975c9bc608d8b62b5987c2}
      \field{sortinit}{C}
      \field{labelyear}{2004}
      \field{labeltitle}{Likelihood ratio tests in linear mixed models with one variance component}
      \field{journaltitle}{Journal of the Royal Statistical Society: Series B (Statistical Methodology)}
      \field{number}{1}
      \field{title}{Likelihood ratio tests in linear mixed models with one variance component}
      \field{volume}{66}
      \field{year}{2004}
      \field{pages}{165\bibrangedash 185}
    \endentry
    \entry{febrero2013gam}{article}{}
      \name{labelname}{3}{}{%
        {{hash=0211aa77511ae639afbb593b0975f920}{Febrero-Bande}{F\bibinithyphendelim B\bibinitperiod}{M.}{M\bibinitperiod}{}{}{}{}}%
        {{hash=5414f627bbbd75ccdc8e827c457c6a81}{Galeano}{G\bibinitperiod}{P.}{P\bibinitperiod}{}{}{}{}}%
        {{hash=9d3ff96ec06c76cdc4ba7bc0f9c23714}{Gonz{\'a}lez-Manteiga}{G\bibinithyphendelim M\bibinitperiod}{W.}{W\bibinitperiod}{}{}{}{}}%
      }
      \name{author}{3}{}{%
        {{hash=0211aa77511ae639afbb593b0975f920}{Febrero-Bande}{F\bibinithyphendelim B\bibinitperiod}{M.}{M\bibinitperiod}{}{}{}{}}%
        {{hash=5414f627bbbd75ccdc8e827c457c6a81}{Galeano}{G\bibinitperiod}{P.}{P\bibinitperiod}{}{}{}{}}%
        {{hash=9d3ff96ec06c76cdc4ba7bc0f9c23714}{Gonz{\'a}lez-Manteiga}{G\bibinithyphendelim M\bibinitperiod}{W.}{W\bibinitperiod}{}{}{}{}}%
      }
      \list{publisher}{1}{%
        {Springer}%
      }
      \strng{namehash}{63325554a3eda3e2c17a914027c0d221}
      \strng{fullhash}{420002b84372bcc6794fba814d38d423}
      \field{sortinit}{F}
      \field{labelyear}{2013}
      \field{labeltitle}{Generalized additive models for functional data}
      \field{journaltitle}{{TEST}}
      \field{number}{2}
      \field{title}{Generalized additive models for functional data}
      \field{volume}{22}
      \field{year}{2013}
      \field{pages}{278\bibrangedash 292}
    \endentry
    \entry{febrero2013package}{article}{}
      \name{labelname}{2}{}{%
        {{hash=0211aa77511ae639afbb593b0975f920}{Febrero-Bande}{F\bibinithyphendelim B\bibinitperiod}{M.}{M\bibinitperiod}{}{}{}{}}%
        {{hash=fc6d005fa19b687c93418da163394800}{{Oviedo\bibnamedelimb de\bibnamedelimb la\bibnamedelimb Fuente}}{O\bibinitperiod}{M.}{M\bibinitperiod}{}{}{}{}}%
      }
      \name{author}{2}{}{%
        {{hash=0211aa77511ae639afbb593b0975f920}{Febrero-Bande}{F\bibinithyphendelim B\bibinitperiod}{M.}{M\bibinitperiod}{}{}{}{}}%
        {{hash=fc6d005fa19b687c93418da163394800}{{Oviedo\bibnamedelimb de\bibnamedelimb la\bibnamedelimb Fuente}}{O\bibinitperiod}{M.}{M\bibinitperiod}{}{}{}{}}%
      }
      \strng{namehash}{eade308668a9c319261c5c0d9da2987a}
      \strng{fullhash}{eade308668a9c319261c5c0d9da2987a}
      \field{sortinit}{F}
      \field{labelyear}{2012}
      \field{labeltitle}{Statistical Computing in Functional Data Analysis: The {R} Package {fda.usc}}
      \field{journaltitle}{Journal of Statistical Software}
      \field{number}{4}
      \field{title}{Statistical Computing in Functional Data Analysis: The {R} Package {fda.usc}}
      \field{urlday}{06}
      \field{urlmonth}{10}
      \field{urlyear}{2013}
      \field{volume}{51}
      \field{year}{2012}
      \field{pages}{1\bibrangedash 28}
      \verb{url}
      \verb http://www.jstatsoft.org/v51/i04/
      \endverb
    \endentry
    \entry{ferraty2006Nonparametric}{book}{}
      \name{labelname}{2}{}{%
        {{hash=e19d2a92dc7a4091d092b2e8b12b52ba}{Ferraty}{F\bibinitperiod}{F.}{F\bibinitperiod}{}{}{}{}}%
        {{hash=919b238b28abef467c8d352e1bd54579}{Vieu}{V\bibinitperiod}{P.}{P\bibinitperiod}{}{}{}{}}%
      }
      \name{author}{2}{}{%
        {{hash=e19d2a92dc7a4091d092b2e8b12b52ba}{Ferraty}{F\bibinitperiod}{F.}{F\bibinitperiod}{}{}{}{}}%
        {{hash=919b238b28abef467c8d352e1bd54579}{Vieu}{V\bibinitperiod}{P.}{P\bibinitperiod}{}{}{}{}}%
      }
      \list{publisher}{1}{%
        {Springer Verlag}%
      }
      \strng{namehash}{ceeeba58f2c2a075319c519cb818f6b8}
      \strng{fullhash}{ceeeba58f2c2a075319c519cb818f6b8}
      \field{sortinit}{F}
      \field{extrayear}{1}
      \field{labelyear}{2006}
      \field{labeltitle}{Nonparametric Functional Data Analysis: Theory and Practice}
      \field{title}{Nonparametric Functional Data Analysis: Theory and Practice}
      \field{year}{2006}
    \endentry
    \entry{staph2006code}{online}{}
      \name{labelname}{2}{}{%
        {{hash=e19d2a92dc7a4091d092b2e8b12b52ba}{Ferraty}{F\bibinitperiod}{F.}{F\bibinitperiod}{}{}{}{}}%
        {{hash=919b238b28abef467c8d352e1bd54579}{Vieu}{V\bibinitperiod}{P.}{P\bibinitperiod}{}{}{}{}}%
      }
      \name{author}{2}{}{%
        {{hash=e19d2a92dc7a4091d092b2e8b12b52ba}{Ferraty}{F\bibinitperiod}{F.}{F\bibinitperiod}{}{}{}{}}%
        {{hash=919b238b28abef467c8d352e1bd54579}{Vieu}{V\bibinitperiod}{P.}{P\bibinitperiod}{}{}{}{}}%
      }
      \strng{namehash}{ceeeba58f2c2a075319c519cb818f6b8}
      \strng{fullhash}{ceeeba58f2c2a075319c519cb818f6b8}
      \field{sortinit}{F}
      \field{extrayear}{2}
      \field{labelyear}{2006}
      \field{labeltitle}{{npfda} {NonParametric} Functional Data Analysis}
      \field{title}{{npfda} {NonParametric} Functional Data Analysis}
      \field{urlday}{08}
      \field{urlmonth}{10}
      \field{urlyear}{2013}
      \field{year}{2006}
      \verb{url}
      \verb http://www.math.univ-toulouse.fr/staph/npfda/
      \endverb
    \endentry
    \entry{gabrys2010tests}{article}{}
      \name{labelname}{3}{}{%
        {{hash=bef514c87e0f735f6b088fd279170e81}{Gabrys}{G\bibinitperiod}{R.}{R\bibinitperiod}{}{}{}{}}%
        {{hash=abd124f6a1c9d14b0259b8025ea6c115}{Horv{\'a}th}{H\bibinitperiod}{L.}{L\bibinitperiod}{}{}{}{}}%
        {{hash=a93529c87a56011d65e4a389e1f3cfdb}{Kokoszka}{K\bibinitperiod}{P.}{P\bibinitperiod}{}{}{}{}}%
      }
      \name{author}{3}{}{%
        {{hash=bef514c87e0f735f6b088fd279170e81}{Gabrys}{G\bibinitperiod}{R.}{R\bibinitperiod}{}{}{}{}}%
        {{hash=abd124f6a1c9d14b0259b8025ea6c115}{Horv{\'a}th}{H\bibinitperiod}{L.}{L\bibinitperiod}{}{}{}{}}%
        {{hash=a93529c87a56011d65e4a389e1f3cfdb}{Kokoszka}{K\bibinitperiod}{P.}{P\bibinitperiod}{}{}{}{}}%
      }
      \list{publisher}{1}{%
        {Taylor \& Francis}%
      }
      \strng{namehash}{091c35ee23708d00d72145bea75d77e2}
      \strng{fullhash}{c987add0c24a74d2a5ce0f2dc250b577}
      \field{sortinit}{G}
      \field{labelyear}{2010}
      \field{labeltitle}{Tests for error correlation in the functional linear model}
      \field{journaltitle}{Journal of the American Statistical Association}
      \field{number}{491}
      \field{title}{Tests for error correlation in the functional linear model}
      \field{volume}{105}
      \field{year}{2010}
      \field{pages}{1113\bibrangedash 1125}
    \endentry
    \entry{febrero2013goodness}{article}{}
      \name{labelname}{3}{}{%
        {{hash=7677232e85b6b8ab85d8cb1a0c260cd6}{Garc{\'i}a-Portugu{\'e}s}{G\bibinithyphendelim P\bibinitperiod}{E}{E\bibinitperiod}{}{}{}{}}%
        {{hash=9d3ff96ec06c76cdc4ba7bc0f9c23714}{Gonz{\'a}lez-Manteiga}{G\bibinithyphendelim M\bibinitperiod}{W.}{W\bibinitperiod}{}{}{}{}}%
        {{hash=0211aa77511ae639afbb593b0975f920}{Febrero-Bande}{F\bibinithyphendelim B\bibinitperiod}{M.}{M\bibinitperiod}{}{}{}{}}%
      }
      \name{author}{3}{}{%
        {{hash=7677232e85b6b8ab85d8cb1a0c260cd6}{Garc{\'i}a-Portugu{\'e}s}{G\bibinithyphendelim P\bibinitperiod}{E}{E\bibinitperiod}{}{}{}{}}%
        {{hash=9d3ff96ec06c76cdc4ba7bc0f9c23714}{Gonz{\'a}lez-Manteiga}{G\bibinithyphendelim M\bibinitperiod}{W.}{W\bibinitperiod}{}{}{}{}}%
        {{hash=0211aa77511ae639afbb593b0975f920}{Febrero-Bande}{F\bibinithyphendelim B\bibinitperiod}{M.}{M\bibinitperiod}{}{}{}{}}%
      }
      \list{publisher}{1}{%
        {Taylor \& Francis}%
      }
      \strng{namehash}{76625f47ca590a028efa3d04eb437e2c}
      \strng{fullhash}{7db41de6157ca44ddfac56a082c1dc97}
      \field{sortinit}{G}
      \field{labelyear}{2013}
      \field{labeltitle}{A goodness-of-fit test for the functional linear model with scalar response}
      \field{journaltitle}{Journal of Computational and Graphical Statistics}
      \field{title}{A goodness-of-fit test for the functional linear model with scalar response}
      \field{year}{2013}
      \field{pages}{to appear}
    \endentry
    \entry{goldsmith2011penalized}{article}{}
      \name{labelname}{5}{}{%
        {{hash=947b47a6cf08079790bb15f78caa2124}{Goldsmith}{G\bibinitperiod}{J.}{J\bibinitperiod}{}{}{}{}}%
        {{hash=d505b6a7a55604884d300bd43c92526f}{Bobb}{B\bibinitperiod}{J.}{J\bibinitperiod}{}{}{}{}}%
        {{hash=6dac3bc879f150e7889487a5f8816d26}{Crainiceanu}{C\bibinitperiod}{C.\bibnamedelimi M.}{C\bibinitperiod\bibinitdelim M\bibinitperiod}{}{}{}{}}%
        {{hash=be361e330970672e21e0887dfafd8b0a}{Caffo}{C\bibinitperiod}{B.}{B\bibinitperiod}{}{}{}{}}%
        {{hash=d08e693e8243dc7c7c1c5e59695dc865}{Reich}{R\bibinitperiod}{D.}{D\bibinitperiod}{}{}{}{}}%
      }
      \name{author}{5}{}{%
        {{hash=947b47a6cf08079790bb15f78caa2124}{Goldsmith}{G\bibinitperiod}{J.}{J\bibinitperiod}{}{}{}{}}%
        {{hash=d505b6a7a55604884d300bd43c92526f}{Bobb}{B\bibinitperiod}{J.}{J\bibinitperiod}{}{}{}{}}%
        {{hash=6dac3bc879f150e7889487a5f8816d26}{Crainiceanu}{C\bibinitperiod}{C.\bibnamedelimi M.}{C\bibinitperiod\bibinitdelim M\bibinitperiod}{}{}{}{}}%
        {{hash=be361e330970672e21e0887dfafd8b0a}{Caffo}{C\bibinitperiod}{B.}{B\bibinitperiod}{}{}{}{}}%
        {{hash=d08e693e8243dc7c7c1c5e59695dc865}{Reich}{R\bibinitperiod}{D.}{D\bibinitperiod}{}{}{}{}}%
      }
      \list{publisher}{1}{%
        {American Statistical Association}%
      }
      \strng{namehash}{bfac0cc5fea2f37a84b625544b18134b}
      \strng{fullhash}{1813d344c7c761581500634351acfd00}
      \field{sortinit}{G}
      \field{labelyear}{2011}
      \field{labeltitle}{Penalized Functional Regression}
      \field{journaltitle}{Journal of Computational and Graphical Statistics}
      \field{number}{4}
      \field{title}{Penalized Functional Regression}
      \field{volume}{20}
      \field{year}{2011}
      \field{pages}{830\bibrangedash 851}
    \endentry
    \entry{greven2008Restricted}{article}{}
      \name{labelname}{4}{}{%
        {{hash=5f02aeb2e784c9b97f0ef51b400929b7}{Greven}{G\bibinitperiod}{S.}{S\bibinitperiod}{}{}{}{}}%
        {{hash=6dac3bc879f150e7889487a5f8816d26}{Crainiceanu}{C\bibinitperiod}{C.\bibnamedelimi M.}{C\bibinitperiod\bibinitdelim M\bibinitperiod}{}{}{}{}}%
        {{hash=ffebf1da16f15efd0a4b4f67ee5635a9}{K{\"u}chenhoff}{K\bibinitperiod}{H.}{H\bibinitperiod}{}{}{}{}}%
        {{hash=996b326321779f31dd8a86be38f9708a}{Peters}{P\bibinitperiod}{A.}{A\bibinitperiod}{}{}{}{}}%
      }
      \name{author}{4}{}{%
        {{hash=5f02aeb2e784c9b97f0ef51b400929b7}{Greven}{G\bibinitperiod}{S.}{S\bibinitperiod}{}{}{}{}}%
        {{hash=6dac3bc879f150e7889487a5f8816d26}{Crainiceanu}{C\bibinitperiod}{C.\bibnamedelimi M.}{C\bibinitperiod\bibinitdelim M\bibinitperiod}{}{}{}{}}%
        {{hash=ffebf1da16f15efd0a4b4f67ee5635a9}{K{\"u}chenhoff}{K\bibinitperiod}{H.}{H\bibinitperiod}{}{}{}{}}%
        {{hash=996b326321779f31dd8a86be38f9708a}{Peters}{P\bibinitperiod}{A.}{A\bibinitperiod}{}{}{}{}}%
      }
      \list{publisher}{1}{%
        {American Statistical Association}%
      }
      \strng{namehash}{f17adee8125f93fd4f1120e0cf0385aa}
      \strng{fullhash}{d07143fb2ab42f5a2b76f3933f3556c0}
      \field{sortinit}{G}
      \field{labelyear}{2008}
      \field{labeltitle}{Restricted likelihood ratio testing for zero variance components in linear mixed models}
      \field{journaltitle}{Journal of Computational and Graphical Statistics}
      \field{number}{4}
      \field{title}{Restricted likelihood ratio testing for zero variance components in linear mixed models}
      \field{volume}{17}
      \field{year}{2008}
      \field{pages}{870\bibrangedash 891}
    \endentry
    \entry{guillas2010bivariate}{article}{}
      \name{labelname}{2}{}{%
        {{hash=8df63a0abcfe31a5b5924e3df0da6977}{Guillas}{G\bibinitperiod}{S.}{S\bibinitperiod}{}{}{}{}}%
        {{hash=b92c64d26d47a29d2f318bd1c0d5ac07}{Lai}{L\bibinitperiod}{M.\bibnamedelimi J.}{M\bibinitperiod\bibinitdelim J\bibinitperiod}{}{}{}{}}%
      }
      \name{author}{2}{}{%
        {{hash=8df63a0abcfe31a5b5924e3df0da6977}{Guillas}{G\bibinitperiod}{S.}{S\bibinitperiod}{}{}{}{}}%
        {{hash=b92c64d26d47a29d2f318bd1c0d5ac07}{Lai}{L\bibinitperiod}{M.\bibnamedelimi J.}{M\bibinitperiod\bibinitdelim J\bibinitperiod}{}{}{}{}}%
      }
      \list{publisher}{1}{%
        {Taylor \& Francis}%
      }
      \strng{namehash}{e3e967e7b978b788fd0dbbabfc17de9e}
      \strng{fullhash}{e3e967e7b978b788fd0dbbabfc17de9e}
      \field{sortinit}{G}
      \field{labelyear}{2010}
      \field{labeltitle}{Bivariate splines for spatial functional regression models}
      \field{journaltitle}{Journal of Nonparametric Statistics}
      \field{number}{4}
      \field{title}{Bivariate splines for spatial functional regression models}
      \field{volume}{22}
      \field{year}{2010}
      \field{pages}{477\bibrangedash 497}
    \endentry
    \entry{horvath2012inference}{book}{}
      \name{labelname}{2}{}{%
        {{hash=abd124f6a1c9d14b0259b8025ea6c115}{Horv{\'a}th}{H\bibinitperiod}{L.}{L\bibinitperiod}{}{}{}{}}%
        {{hash=a93529c87a56011d65e4a389e1f3cfdb}{Kokoszka}{K\bibinitperiod}{P.}{P\bibinitperiod}{}{}{}{}}%
      }
      \name{author}{2}{}{%
        {{hash=abd124f6a1c9d14b0259b8025ea6c115}{Horv{\'a}th}{H\bibinitperiod}{L.}{L\bibinitperiod}{}{}{}{}}%
        {{hash=a93529c87a56011d65e4a389e1f3cfdb}{Kokoszka}{K\bibinitperiod}{P.}{P\bibinitperiod}{}{}{}{}}%
      }
      \list{publisher}{1}{%
        {Springer}%
      }
      \strng{namehash}{bf94b9a32e4af4e0123232cab9abff1a}
      \strng{fullhash}{bf94b9a32e4af4e0123232cab9abff1a}
      \field{sortinit}{H}
      \field{labelyear}{2012}
      \field{labeltitle}{Inference for functional data with applications}
      \field{title}{Inference for functional data with applications}
      \field{year}{2012}
    \endentry
    \entry{james2005functional}{article}{}
      \name{labelname}{2}{}{%
        {{hash=89476f0d8069e4406ab3664d210d5bca}{James}{J\bibinitperiod}{G.\bibnamedelimi M.}{G\bibinitperiod\bibinitdelim M\bibinitperiod}{}{}{}{}}%
        {{hash=1a1448224e288d36850d614d839ef85c}{Silverman}{S\bibinitperiod}{B.\bibnamedelimi W.}{B\bibinitperiod\bibinitdelim W\bibinitperiod}{}{}{}{}}%
      }
      \name{author}{2}{}{%
        {{hash=89476f0d8069e4406ab3664d210d5bca}{James}{J\bibinitperiod}{G.\bibnamedelimi M.}{G\bibinitperiod\bibinitdelim M\bibinitperiod}{}{}{}{}}%
        {{hash=1a1448224e288d36850d614d839ef85c}{Silverman}{S\bibinitperiod}{B.\bibnamedelimi W.}{B\bibinitperiod\bibinitdelim W\bibinitperiod}{}{}{}{}}%
      }
      \list{publisher}{1}{%
        {American Statistical Association}%
      }
      \strng{namehash}{78805381252c28670189e7de973f337b}
      \strng{fullhash}{78805381252c28670189e7de973f337b}
      \field{sortinit}{J}
      \field{labelyear}{2005}
      \field{labeltitle}{Functional Adaptive Model Estimation.}
      \field{journaltitle}{Journal of the American Statistical Association}
      \field{number}{470}
      \field{title}{Functional Adaptive Model Estimation.}
      \field{volume}{100}
      \field{year}{2005}
      \field{pages}{565\bibrangedash 577}
    \endentry
    \entry{staicu2013classical}{online}{}
      \name{labelname}{3}{}{%
        {{hash=39c513a7a12603c47cbd47ed84ac1114}{Kong}{K\bibinitperiod}{D.}{D\bibinitperiod}{}{}{}{}}%
        {{hash=af9625935a7c7ff3f36dd8530f62b188}{Staicu}{S\bibinitperiod}{A.\bibnamedelimi M.}{A\bibinitperiod\bibinitdelim M\bibinitperiod}{}{}{}{}}%
        {{hash=427c1b4a3abedc9a0a08fcce6369cce8}{Maity}{M\bibinitperiod}{A.}{A\bibinitperiod}{}{}{}{}}%
      }
      \name{author}{3}{}{%
        {{hash=39c513a7a12603c47cbd47ed84ac1114}{Kong}{K\bibinitperiod}{D.}{D\bibinitperiod}{}{}{}{}}%
        {{hash=af9625935a7c7ff3f36dd8530f62b188}{Staicu}{S\bibinitperiod}{A.\bibnamedelimi M.}{A\bibinitperiod\bibinitdelim M\bibinitperiod}{}{}{}{}}%
        {{hash=427c1b4a3abedc9a0a08fcce6369cce8}{Maity}{M\bibinitperiod}{A.}{A\bibinitperiod}{}{}{}{}}%
      }
      \strng{namehash}{9dbd6757b8f46a18477708dc00e70f79}
      \strng{fullhash}{f0d1580716ea268d36e8bf653ff1fd74}
      \field{sortinit}{K}
      \field{labelyear}{2013}
      \field{labeltitle}{Classical Testing in Functional Linear Models}
      \field{pubstate}{Submitted}
      \field{title}{Classical Testing in Functional Linear Models}
      \field{urlday}{06}
      \field{urlmonth}{10}
      \field{urlyear}{2013}
      \field{year}{2013}
      \verb{url}
      \verb http://www4.stat.ncsu.edu/~staicu/papers/ClassicalTest_FLM_KSM.pdf
      \endverb
    \endentry
    \entry{li2010generalized}{article}{}
      \name{labelname}{3}{}{%
        {{hash=38fcf426aad9f6b84953f797dca3a801}{Li}{L\bibinitperiod}{Y.}{Y\bibinitperiod}{}{}{}{}}%
        {{hash=117fb2be4b6fdec3f9c499da50257905}{Wang}{W\bibinitperiod}{N.}{N\bibinitperiod}{}{}{}{}}%
        {{hash=10110f837cefb25fa3101225492ef3ce}{Carroll}{C\bibinitperiod}{R.\bibnamedelimi J.}{R\bibinitperiod\bibinitdelim J\bibinitperiod}{}{}{}{}}%
      }
      \name{author}{3}{}{%
        {{hash=38fcf426aad9f6b84953f797dca3a801}{Li}{L\bibinitperiod}{Y.}{Y\bibinitperiod}{}{}{}{}}%
        {{hash=117fb2be4b6fdec3f9c499da50257905}{Wang}{W\bibinitperiod}{N.}{N\bibinitperiod}{}{}{}{}}%
        {{hash=10110f837cefb25fa3101225492ef3ce}{Carroll}{C\bibinitperiod}{R.\bibnamedelimi J.}{R\bibinitperiod\bibinitdelim J\bibinitperiod}{}{}{}{}}%
      }
      \strng{namehash}{cc196a5ce1f3c4120fc08f1048e1df69}
      \strng{fullhash}{d5987c242a9167fdc305a0389ee97d21}
      \field{sortinit}{L}
      \field{labelyear}{2010}
      \field{labeltitle}{Generalized functional linear models with semiparametric single-index interactions}
      \field{journaltitle}{Journal of the American Statistical Association}
      \field{number}{490}
      \field{title}{Generalized functional linear models with semiparametric single-index interactions}
      \field{volume}{105}
      \field{year}{2010}
    \endentry
    \entry{marx2005multidimensional}{article}{}
      \name{labelname}{2}{}{%
        {{hash=3c1cd900a81c0ff0648d077268c56880}{Marx}{M\bibinitperiod}{B.\bibnamedelimi D.}{B\bibinitperiod\bibinitdelim D\bibinitperiod}{}{}{}{}}%
        {{hash=90b6cff841b1ea11b2a914024713cd04}{Eilers}{E\bibinitperiod}{P.\bibnamedelimi H.\bibnamedelimi C.}{P\bibinitperiod\bibinitdelim H\bibinitperiod\bibinitdelim C\bibinitperiod}{}{}{}{}}%
      }
      \name{author}{2}{}{%
        {{hash=3c1cd900a81c0ff0648d077268c56880}{Marx}{M\bibinitperiod}{B.\bibnamedelimi D.}{B\bibinitperiod\bibinitdelim D\bibinitperiod}{}{}{}{}}%
        {{hash=90b6cff841b1ea11b2a914024713cd04}{Eilers}{E\bibinitperiod}{P.\bibnamedelimi H.\bibnamedelimi C.}{P\bibinitperiod\bibinitdelim H\bibinitperiod\bibinitdelim C\bibinitperiod}{}{}{}{}}%
      }
      \list{publisher}{1}{%
        {ASA}%
      }
      \strng{namehash}{104c9d24a97a070449155ae532cf6b1d}
      \strng{fullhash}{104c9d24a97a070449155ae532cf6b1d}
      \field{sortinit}{M}
      \field{labelyear}{2005}
      \field{labeltitle}{Multidimensional penalized signal regression}
      \field{journaltitle}{Technometrics}
      \field{number}{1}
      \field{title}{Multidimensional penalized signal regression}
      \field{volume}{47}
      \field{year}{2005}
      \field{pages}{13\bibrangedash 22}
    \endentry
    \entry{McLean2012functional}{article}{}
      \name{labelname}{5}{}{%
        {{hash=ddb82a9e327b7966f04124ab19540f69}{McLean}{M\bibinitperiod}{M.\bibnamedelimi W.}{M\bibinitperiod\bibinitdelim W\bibinitperiod}{}{}{}{}}%
        {{hash=b4534ebe5315666e28756f054a7a003c}{Hooker}{H\bibinitperiod}{G.}{G\bibinitperiod}{}{}{}{}}%
        {{hash=af9625935a7c7ff3f36dd8530f62b188}{Staicu}{S\bibinitperiod}{A.\bibnamedelimi M.}{A\bibinitperiod\bibinitdelim M\bibinitperiod}{}{}{}{}}%
        {{hash=4f1e5e96ac9e3d1124e4f0be96e8c9f6}{Scheipl}{S\bibinitperiod}{F.}{F\bibinitperiod}{}{}{}{}}%
        {{hash=6c4cb8ccb3c3853c45b3e9179df4dea2}{Ruppert}{R\bibinitperiod}{D.}{D\bibinitperiod}{}{}{}{}}%
      }
      \name{author}{5}{}{%
        {{hash=ddb82a9e327b7966f04124ab19540f69}{McLean}{M\bibinitperiod}{M.\bibnamedelimi W.}{M\bibinitperiod\bibinitdelim W\bibinitperiod}{}{}{}{}}%
        {{hash=b4534ebe5315666e28756f054a7a003c}{Hooker}{H\bibinitperiod}{G.}{G\bibinitperiod}{}{}{}{}}%
        {{hash=af9625935a7c7ff3f36dd8530f62b188}{Staicu}{S\bibinitperiod}{A.\bibnamedelimi M.}{A\bibinitperiod\bibinitdelim M\bibinitperiod}{}{}{}{}}%
        {{hash=4f1e5e96ac9e3d1124e4f0be96e8c9f6}{Scheipl}{S\bibinitperiod}{F.}{F\bibinitperiod}{}{}{}{}}%
        {{hash=6c4cb8ccb3c3853c45b3e9179df4dea2}{Ruppert}{R\bibinitperiod}{D.}{D\bibinitperiod}{}{}{}{}}%
      }
      \strng{namehash}{bf9421d2d4db929001fc2a5df4d51e09}
      \strng{fullhash}{e216c10884d0dd7d699030035e6eb9cf}
      \field{sortinit}{M}
      \field{extrayear}{1}
      \field{labelyear}{2013}
      \field{labeltitle}{Functional Generalized Additive Models}
      \field{journaltitle}{Journal of Computational and Graphical Statistics}
      \field{title}{Functional Generalized Additive Models}
      \field{urlday}{06}
      \field{urlmonth}{10}
      \field{urlyear}{2013}
      \field{year}{2013}
      \verb{doi}
      \verb 10.1080/10618600.2012.729985
      \endverb
      \verb{url}
      \verb http://amstat.tandfonline.com/doi/full/10.1080/10618600.2012.729985
      \endverb
    \endentry
    \entry{mclean2013bayesian}{online}{}
      \name{labelname}{5}{}{%
        {{hash=ddb82a9e327b7966f04124ab19540f69}{McLean}{M\bibinitperiod}{M.\bibnamedelimi W.}{M\bibinitperiod\bibinitdelim W\bibinitperiod}{}{}{}{}}%
        {{hash=4f1e5e96ac9e3d1124e4f0be96e8c9f6}{Scheipl}{S\bibinitperiod}{F.}{F\bibinitperiod}{}{}{}{}}%
        {{hash=b4534ebe5315666e28756f054a7a003c}{Hooker}{H\bibinitperiod}{G.}{G\bibinitperiod}{}{}{}{}}%
        {{hash=5f02aeb2e784c9b97f0ef51b400929b7}{Greven}{G\bibinitperiod}{S.}{S\bibinitperiod}{}{}{}{}}%
        {{hash=6c4cb8ccb3c3853c45b3e9179df4dea2}{Ruppert}{R\bibinitperiod}{D.}{D\bibinitperiod}{}{}{}{}}%
      }
      \name{author}{5}{}{%
        {{hash=ddb82a9e327b7966f04124ab19540f69}{McLean}{M\bibinitperiod}{M.\bibnamedelimi W.}{M\bibinitperiod\bibinitdelim W\bibinitperiod}{}{}{}{}}%
        {{hash=4f1e5e96ac9e3d1124e4f0be96e8c9f6}{Scheipl}{S\bibinitperiod}{F.}{F\bibinitperiod}{}{}{}{}}%
        {{hash=b4534ebe5315666e28756f054a7a003c}{Hooker}{H\bibinitperiod}{G.}{G\bibinitperiod}{}{}{}{}}%
        {{hash=5f02aeb2e784c9b97f0ef51b400929b7}{Greven}{G\bibinitperiod}{S.}{S\bibinitperiod}{}{}{}{}}%
        {{hash=6c4cb8ccb3c3853c45b3e9179df4dea2}{Ruppert}{R\bibinitperiod}{D.}{D\bibinitperiod}{}{}{}{}}%
      }
      \strng{namehash}{bf9421d2d4db929001fc2a5df4d51e09}
      \strng{fullhash}{b68147e1203ef1b60fee421c77ab7c40}
      \field{sortinit}{M}
      \field{extrayear}{2}
      \field{labelyear}{2013}
      \field{labeltitle}{Bayesian Functional Generalized Additive Models with Sparsely Observed Covariates}
      \field{eprintclass}{stat.ME}
      \field{eprinttype}{arxiv}
      \field{title}{Bayesian Functional Generalized Additive Models with Sparsely Observed Covariates}
      \field{urlday}{06}
      \field{urlmonth}{10}
      \field{urlyear}{2013}
      \field{year}{2013}
      \verb{eprint}
      \verb 1305.3585
      \endverb
    \endentry
    \entry{mueller2013continuously}{article}{}
      \name{labelname}{3}{}{%
        {{hash=26ae59136e82514e73fdff084edf19c7}{M{\"u}ller}{M\bibinitperiod}{H.\bibnamedelimi G.}{H\bibinitperiod\bibinitdelim G\bibinitperiod}{}{}{}{}}%
        {{hash=57cfccefdf02c4242604fbd44d33d9e1}{Wu}{W\bibinitperiod}{Y.}{Y\bibinitperiod}{}{}{}{}}%
        {{hash=072b813073313559f9c145e9c63501f7}{Yao}{Y\bibinitperiod}{F.}{F\bibinitperiod}{}{}{}{}}%
      }
      \name{author}{3}{}{%
        {{hash=26ae59136e82514e73fdff084edf19c7}{M{\"u}ller}{M\bibinitperiod}{H.\bibnamedelimi G.}{H\bibinitperiod\bibinitdelim G\bibinitperiod}{}{}{}{}}%
        {{hash=57cfccefdf02c4242604fbd44d33d9e1}{Wu}{W\bibinitperiod}{Y.}{Y\bibinitperiod}{}{}{}{}}%
        {{hash=072b813073313559f9c145e9c63501f7}{Yao}{Y\bibinitperiod}{F.}{F\bibinitperiod}{}{}{}{}}%
      }
      \strng{namehash}{4cf19a5663b8935a48f5476193ae3023}
      \strng{fullhash}{3317243c43b0a069e50e7cebfd35c532}
      \field{sortinit}{M}
      \field{labelyear}{2013}
      \field{labeltitle}{Continuously additive models for nonlinear functional regression}
      \field{journaltitle}{Biometrika}
      \field{number}{3}
      \field{title}{Continuously additive models for nonlinear functional regression}
      \field{urlday}{06}
      \field{urlmonth}{10}
      \field{urlyear}{2013}
      \field{volume}{100}
      \field{year}{2013}
      \verb{doi}
      \verb 10.1093/biomet/ast004
      \endverb
      \verb{url}
      \verb http://biomet.oxfordjournals.org/content/100/3/607
      \endverb
    \endentry
    \entry{muller2008functional}{article}{}
      \name{labelname}{2}{}{%
        {{hash=26ae59136e82514e73fdff084edf19c7}{M{\"u}ller}{M\bibinitperiod}{H.\bibnamedelimi G.}{H\bibinitperiod\bibinitdelim G\bibinitperiod}{}{}{}{}}%
        {{hash=072b813073313559f9c145e9c63501f7}{Yao}{Y\bibinitperiod}{F.}{F\bibinitperiod}{}{}{}{}}%
      }
      \name{author}{2}{}{%
        {{hash=26ae59136e82514e73fdff084edf19c7}{M{\"u}ller}{M\bibinitperiod}{H.\bibnamedelimi G.}{H\bibinitperiod\bibinitdelim G\bibinitperiod}{}{}{}{}}%
        {{hash=072b813073313559f9c145e9c63501f7}{Yao}{Y\bibinitperiod}{F.}{F\bibinitperiod}{}{}{}{}}%
      }
      \list{publisher}{1}{%
        {ASA}%
      }
      \strng{namehash}{a951db33c8caf87102df4bec45959fbe}
      \strng{fullhash}{a951db33c8caf87102df4bec45959fbe}
      \field{sortinit}{M}
      \field{labelyear}{2008}
      \field{labeltitle}{Functional additive models}
      \field{issn}{0162-1459}
      \field{journaltitle}{Journal of the American Statistical Association}
      \field{number}{484}
      \field{title}{Functional additive models}
      \field{volume}{103}
      \field{year}{2008}
      \field{pages}{1534\bibrangedash 1544}
    \endentry
    \entry{patterson1971recovery}{article}{}
      \name{labelname}{2}{}{%
        {{hash=732bea2ef63073acf6555fe8f3058166}{Patterson}{P\bibinitperiod}{H.\bibnamedelimi D.}{H\bibinitperiod\bibinitdelim D\bibinitperiod}{}{}{}{}}%
        {{hash=8582a6e39f8ff93d11161c3bc0d04569}{Thompson}{T\bibinitperiod}{R.}{R\bibinitperiod}{}{}{}{}}%
      }
      \name{author}{2}{}{%
        {{hash=732bea2ef63073acf6555fe8f3058166}{Patterson}{P\bibinitperiod}{H.\bibnamedelimi D.}{H\bibinitperiod\bibinitdelim D\bibinitperiod}{}{}{}{}}%
        {{hash=8582a6e39f8ff93d11161c3bc0d04569}{Thompson}{T\bibinitperiod}{R.}{R\bibinitperiod}{}{}{}{}}%
      }
      \list{publisher}{1}{%
        {Biometrika Trust}%
      }
      \strng{namehash}{82429df92ea86d6ce3e80c1947a85767}
      \strng{fullhash}{82429df92ea86d6ce3e80c1947a85767}
      \field{sortinit}{P}
      \field{labelyear}{1971}
      \field{labeltitle}{Recovery of inter-block information when block sizes are unequal}
      \field{journaltitle}{Biometrika}
      \field{number}{3}
      \field{title}{Recovery of inter-block information when block sizes are unequal}
      \field{volume}{58}
      \field{year}{1971}
      \field{pages}{545\bibrangedash 554}
    \endentry
    \entry{pinheiro2000linear}{book}{}
      \name{labelname}{2}{}{%
        {{hash=23c2f8e47ed228f7f26b4e7cd8dafe06}{Pinheiro}{P\bibinitperiod}{J.\bibnamedelimi C.}{J\bibinitperiod\bibinitdelim C\bibinitperiod}{}{}{}{}}%
        {{hash=3f4a487ade2b84dce91e3cc4dcfee97c}{Bates}{B\bibinitperiod}{D.\bibnamedelimi M.}{D\bibinitperiod\bibinitdelim M\bibinitperiod}{}{}{}{}}%
      }
      \name{author}{2}{}{%
        {{hash=23c2f8e47ed228f7f26b4e7cd8dafe06}{Pinheiro}{P\bibinitperiod}{J.\bibnamedelimi C.}{J\bibinitperiod\bibinitdelim C\bibinitperiod}{}{}{}{}}%
        {{hash=3f4a487ade2b84dce91e3cc4dcfee97c}{Bates}{B\bibinitperiod}{D.\bibnamedelimi M.}{D\bibinitperiod\bibinitdelim M\bibinitperiod}{}{}{}{}}%
      }
      \list{publisher}{1}{%
        {Springer}%
      }
      \strng{namehash}{e29f5ac833f31b842331d3778a44a531}
      \strng{fullhash}{e29f5ac833f31b842331d3778a44a531}
      \field{sortinit}{P}
      \field{labelyear}{2000}
      \field{labeltitle}{Linear mixed-effects models: basic concepts and examples}
      \field{title}{Linear mixed-effects models: basic concepts and examples}
      \field{year}{2000}
    \endentry
    \entry{pinheiro2013nlme}{manual}{}
      \name{labelname}{5}{}{%
        {{hash=23c2f8e47ed228f7f26b4e7cd8dafe06}{Pinheiro}{P\bibinitperiod}{J.\bibnamedelimi C.}{J\bibinitperiod\bibinitdelim C\bibinitperiod}{}{}{}{}}%
        {{hash=3f4a487ade2b84dce91e3cc4dcfee97c}{Bates}{B\bibinitperiod}{D.\bibnamedelimi M.}{D\bibinitperiod\bibinitdelim M\bibinitperiod}{}{}{}{}}%
        {{hash=8f33b56bf13bcdc93c9eb978b9591f6b}{DebRoy}{D\bibinitperiod}{S.}{S\bibinitperiod}{}{}{}{}}%
        {{hash=c66905687b4fa2d7cfd34a81a6077915}{Sarkar}{S\bibinitperiod}{Deepayan}{D\bibinitperiod}{}{}{}{}}%
        {{hash=a902f7a1275f47f57554b7ea75ccd0bc}{{R\bibnamedelimb Core\bibnamedelimb Team}}{R\bibinitperiod}{}{}{}{}{}{}}%
      }
      \name{author}{5}{}{%
        {{hash=23c2f8e47ed228f7f26b4e7cd8dafe06}{Pinheiro}{P\bibinitperiod}{J.\bibnamedelimi C.}{J\bibinitperiod\bibinitdelim C\bibinitperiod}{}{}{}{}}%
        {{hash=3f4a487ade2b84dce91e3cc4dcfee97c}{Bates}{B\bibinitperiod}{D.\bibnamedelimi M.}{D\bibinitperiod\bibinitdelim M\bibinitperiod}{}{}{}{}}%
        {{hash=8f33b56bf13bcdc93c9eb978b9591f6b}{DebRoy}{D\bibinitperiod}{S.}{S\bibinitperiod}{}{}{}{}}%
        {{hash=c66905687b4fa2d7cfd34a81a6077915}{Sarkar}{S\bibinitperiod}{Deepayan}{D\bibinitperiod}{}{}{}{}}%
        {{hash=a902f7a1275f47f57554b7ea75ccd0bc}{{R\bibnamedelimb Core\bibnamedelimb Team}}{R\bibinitperiod}{}{}{}{}{}{}}%
      }
      \strng{namehash}{c086cb720348e3211fabf16d5a231ec9}
      \strng{fullhash}{88493a16e6f487488fe4ce9db38273a0}
      \field{sortinit}{P}
      \field{labelyear}{2013}
      \field{labeltitle}{{nlme}: Linear and Nonlinear Mixed Effects Models}
      \field{note}{R package version 3.1-111}
      \field{title}{{nlme}: Linear and Nonlinear Mixed Effects Models}
      \field{year}{2013}
    \endentry
    \entry{r2012}{manual}{}
      \name{labelname}{1}{}{%
        {{hash=a902f7a1275f47f57554b7ea75ccd0bc}{{R\bibnamedelimb Core\bibnamedelimb Team}}{R\bibinitperiod}{}{}{}{}{}{}}%
      }
      \name{author}{1}{}{%
        {{hash=a902f7a1275f47f57554b7ea75ccd0bc}{{R\bibnamedelimb Core\bibnamedelimb Team}}{R\bibinitperiod}{}{}{}{}{}{}}%
      }
      \list{location}{1}{%
        {Vienna, Austria}%
      }
      \list{organization}{1}{%
        {R Foundation for Statistical Computing}%
      }
      \strng{namehash}{a902f7a1275f47f57554b7ea75ccd0bc}
      \strng{fullhash}{a902f7a1275f47f57554b7ea75ccd0bc}
      \field{sortinit}{R}
      \field{labelyear}{2012}
      \field{labeltitle}{R: A Language and Environment for Statistical Computing}
      \field{isbn}{3--900051--07--0}
      \field{title}{R: A Language and Environment for Statistical Computing}
      \field{urlday}{06}
      \field{urlmonth}{10}
      \field{urlyear}{2013}
      \field{year}{2012}
      \verb{url}
      \verb http://www.R-project.org/
      \endverb
    \endentry
    \entry{ramsay2005Functional}{book}{}
      \name{labelname}{2}{}{%
        {{hash=d36d31f91fcb86c921168f2ca2acd908}{Ramsay}{R\bibinitperiod}{J.\bibnamedelimi O.}{J\bibinitperiod\bibinitdelim O\bibinitperiod}{}{}{}{}}%
        {{hash=1a1448224e288d36850d614d839ef85c}{Silverman}{S\bibinitperiod}{B.\bibnamedelimi W.}{B\bibinitperiod\bibinitdelim W\bibinitperiod}{}{}{}{}}%
      }
      \name{author}{2}{}{%
        {{hash=d36d31f91fcb86c921168f2ca2acd908}{Ramsay}{R\bibinitperiod}{J.\bibnamedelimi O.}{J\bibinitperiod\bibinitdelim O\bibinitperiod}{}{}{}{}}%
        {{hash=1a1448224e288d36850d614d839ef85c}{Silverman}{S\bibinitperiod}{B.\bibnamedelimi W.}{B\bibinitperiod\bibinitdelim W\bibinitperiod}{}{}{}{}}%
      }
      \list{publisher}{1}{%
        {Springer}%
      }
      \strng{namehash}{dde46d978244fb05614a830ce2fdfa2a}
      \strng{fullhash}{dde46d978244fb05614a830ce2fdfa2a}
      \field{sortinit}{R}
      \field{labelyear}{2005}
      \field{labeltitle}{Functional Data Analysis}
      \field{edition}{Second}
      \field{title}{Functional Data Analysis}
      \field{year}{2005}
    \endentry
    \entry{reiss2009smoothing}{article}{}
      \name{labelname}{2}{}{%
        {{hash=d888d5b7a3ceef7e6f4be790a8d2a000}{Reiss}{R\bibinitperiod}{P.\bibnamedelimi T.}{P\bibinitperiod\bibinitdelim T\bibinitperiod}{}{}{}{}}%
        {{hash=bb884fe8e77ad17561f9a4b24e3e2dbb}{Ogden}{O\bibinitperiod}{R.\bibnamedelimi T.}{R\bibinitperiod\bibinitdelim T\bibinitperiod}{}{}{}{}}%
      }
      \name{author}{2}{}{%
        {{hash=d888d5b7a3ceef7e6f4be790a8d2a000}{Reiss}{R\bibinitperiod}{P.\bibnamedelimi T.}{P\bibinitperiod\bibinitdelim T\bibinitperiod}{}{}{}{}}%
        {{hash=bb884fe8e77ad17561f9a4b24e3e2dbb}{Ogden}{O\bibinitperiod}{R.\bibnamedelimi T.}{R\bibinitperiod\bibinitdelim T\bibinitperiod}{}{}{}{}}%
      }
      \list{publisher}{1}{%
        {Wiley Online Library}%
      }
      \strng{namehash}{2156eabfa61ac8fb2503db4a44de959b}
      \strng{fullhash}{2156eabfa61ac8fb2503db4a44de959b}
      \field{sortinit}{R}
      \field{labelyear}{2009}
      \field{labeltitle}{Smoothing parameter selection for a class of semiparametric linear models}
      \field{journaltitle}{Journal of the Royal Statistical Society: Series B (Statistical Methodology)}
      \field{number}{2}
      \field{title}{Smoothing parameter selection for a class of semiparametric linear models}
      \field{volume}{71}
      \field{year}{2009}
      \field{pages}{505\bibrangedash 523}
    \endentry
    \entry{ruppert2003semiparametric}{book}{}
      \name{labelname}{3}{}{%
        {{hash=6c4cb8ccb3c3853c45b3e9179df4dea2}{Ruppert}{R\bibinitperiod}{D.}{D\bibinitperiod}{}{}{}{}}%
        {{hash=1ec9f96079fa3006c9492752c62108d4}{Wand}{W\bibinitperiod}{M.\bibnamedelimi P.}{M\bibinitperiod\bibinitdelim P\bibinitperiod}{}{}{}{}}%
        {{hash=10110f837cefb25fa3101225492ef3ce}{Carroll}{C\bibinitperiod}{R.\bibnamedelimi J.}{R\bibinitperiod\bibinitdelim J\bibinitperiod}{}{}{}{}}%
      }
      \name{author}{3}{}{%
        {{hash=6c4cb8ccb3c3853c45b3e9179df4dea2}{Ruppert}{R\bibinitperiod}{D.}{D\bibinitperiod}{}{}{}{}}%
        {{hash=1ec9f96079fa3006c9492752c62108d4}{Wand}{W\bibinitperiod}{M.\bibnamedelimi P.}{M\bibinitperiod\bibinitdelim P\bibinitperiod}{}{}{}{}}%
        {{hash=10110f837cefb25fa3101225492ef3ce}{Carroll}{C\bibinitperiod}{R.\bibnamedelimi J.}{R\bibinitperiod\bibinitdelim J\bibinitperiod}{}{}{}{}}%
      }
      \list{publisher}{1}{%
        {Cambridge University Press}%
      }
      \strng{namehash}{5572406b43d5de8932d98bf4db25ccf6}
      \strng{fullhash}{503411b484bda843ad2862b3a3d6da22}
      \field{sortinit}{R}
      \field{labelyear}{2003}
      \field{labeltitle}{Semiparametric regression}
      \field{title}{Semiparametric regression}
      \field{year}{2003}
    \endentry
    \entry{procmixed2012}{manual}{}
      \name{labelname}{1}{}{%
        {{hash=bd872fab07fa8576cca60ad225e18a4f}{{SAS\bibnamedelimb Institute\bibnamedelimb Inc.}}{S\bibinitperiod}{}{}{}{}{}{}}%
      }
      \name{author}{1}{}{%
        {{hash=bd872fab07fa8576cca60ad225e18a4f}{{SAS\bibnamedelimb Institute\bibnamedelimb Inc.}}{S\bibinitperiod}{}{}{}{}{}{}}%
      }
      \list{location}{1}{%
        {Cary, NC: {SAS} Institute Inc.}%
      }
      \strng{namehash}{bd872fab07fa8576cca60ad225e18a4f}
      \strng{fullhash}{bd872fab07fa8576cca60ad225e18a4f}
      \field{sortinit}{S}
      \field{labelyear}{2008}
      \field{labeltitle}{{SAS/STAT}\textregistered 9.2 User's Guide}
      \field{title}{{SAS/STAT}\textregistered 9.2 User's Guide}
      \field{year}{2008}
    \endentry
    \entry{scheipl2008Size}{article}{}
      \name{labelname}{3}{}{%
        {{hash=4f1e5e96ac9e3d1124e4f0be96e8c9f6}{Scheipl}{S\bibinitperiod}{F.}{F\bibinitperiod}{}{}{}{}}%
        {{hash=5f02aeb2e784c9b97f0ef51b400929b7}{Greven}{G\bibinitperiod}{S.}{S\bibinitperiod}{}{}{}{}}%
        {{hash=ffebf1da16f15efd0a4b4f67ee5635a9}{K{\"u}chenhoff}{K\bibinitperiod}{H.}{H\bibinitperiod}{}{}{}{}}%
      }
      \name{author}{3}{}{%
        {{hash=4f1e5e96ac9e3d1124e4f0be96e8c9f6}{Scheipl}{S\bibinitperiod}{F.}{F\bibinitperiod}{}{}{}{}}%
        {{hash=5f02aeb2e784c9b97f0ef51b400929b7}{Greven}{G\bibinitperiod}{S.}{S\bibinitperiod}{}{}{}{}}%
        {{hash=ffebf1da16f15efd0a4b4f67ee5635a9}{K{\"u}chenhoff}{K\bibinitperiod}{H.}{H\bibinitperiod}{}{}{}{}}%
      }
      \list{publisher}{1}{%
        {Elsevier}%
      }
      \strng{namehash}{28a9369cb28c4011d9b566225dc51d53}
      \strng{fullhash}{aebf47252092d0c81eaac8e2c0d92d72}
      \field{sortinit}{S}
      \field{labelyear}{2008}
      \field{labeltitle}{Size and power of tests for a zero random effect variance or polynomial regression in additive and linear mixed models}
      \field{journaltitle}{Computational Statistics \& Data Analysis}
      \field{number}{7}
      \field{title}{Size and power of tests for a zero random effect variance or polynomial regression in additive and linear mixed models}
      \field{volume}{52}
      \field{year}{2008}
      \field{pages}{3283\bibrangedash 3299}
    \endentry
    \entry{swihart2013restricted}{report}{}
      \name{labelname}{3}{}{%
        {{hash=b2da00662f8a26a443407a63e9d76916}{Swihart}{S\bibinitperiod}{B.\bibnamedelimi J.}{B\bibinitperiod\bibinitdelim J\bibinitperiod}{}{}{}{}}%
        {{hash=947b47a6cf08079790bb15f78caa2124}{Goldsmith}{G\bibinitperiod}{J.}{J\bibinitperiod}{}{}{}{}}%
        {{hash=6dac3bc879f150e7889487a5f8816d26}{Crainiceanu}{C\bibinitperiod}{C.\bibnamedelimi M.}{C\bibinitperiod\bibinitdelim M\bibinitperiod}{}{}{}{}}%
      }
      \name{author}{3}{}{%
        {{hash=b2da00662f8a26a443407a63e9d76916}{Swihart}{S\bibinitperiod}{B.\bibnamedelimi J.}{B\bibinitperiod\bibinitdelim J\bibinitperiod}{}{}{}{}}%
        {{hash=947b47a6cf08079790bb15f78caa2124}{Goldsmith}{G\bibinitperiod}{J.}{J\bibinitperiod}{}{}{}{}}%
        {{hash=6dac3bc879f150e7889487a5f8816d26}{Crainiceanu}{C\bibinitperiod}{C.\bibnamedelimi M.}{C\bibinitperiod\bibinitdelim M\bibinitperiod}{}{}{}{}}%
      }
      \list{institution}{1}{%
        {Johns Hopkins University, Dept. of Biostatistics}%
      }
      \strng{namehash}{f91ecdb7ca2e0f07701269b5ae1cb62c}
      \strng{fullhash}{b8e84aece8aaf35e4953fe20e596ca4f}
      \field{sortinit}{S}
      \field{labelyear}{2013}
      \field{labeltitle}{Restricted likelihood ratio tests for functional effects in the functional linear model}
      \field{number}{247}
      \field{title}{Restricted likelihood ratio tests for functional effects in the functional linear model}
      \field{type}{techreport}
      \field{urlday}{06}
      \field{urlmonth}{10}
      \field{urlyear}{2013}
      \field{year}{2013}
      \verb{url}
      \verb http://biostats.bepress.com/jhubiostat/paper247
      \endverb
    \endentry
    \entry{wang2013optimal}{online}{}
      \name{labelname}{2}{}{%
        {{hash=298db7fe95de35f7ad8f9a6126e3159b}{Wang}{W\bibinitperiod}{X.}{X\bibinitperiod}{}{}{}{}}%
        {{hash=6c4cb8ccb3c3853c45b3e9179df4dea2}{Ruppert}{R\bibinitperiod}{D.}{D\bibinitperiod}{}{}{}{}}%
      }
      \name{author}{2}{}{%
        {{hash=298db7fe95de35f7ad8f9a6126e3159b}{Wang}{W\bibinitperiod}{X.}{X\bibinitperiod}{}{}{}{}}%
        {{hash=6c4cb8ccb3c3853c45b3e9179df4dea2}{Ruppert}{R\bibinitperiod}{D.}{D\bibinitperiod}{}{}{}{}}%
      }
      \strng{namehash}{b268f4bd14419a5a75a38a9a23d10083}
      \strng{fullhash}{b268f4bd14419a5a75a38a9a23d10083}
      \field{sortinit}{W}
      \field{labelyear}{2013}
      \field{labeltitle}{Optimal Prediction in an Additive Functional Model}
      \field{eprintclass}{math.ST}
      \field{eprinttype}{arxiv}
      \field{title}{Optimal Prediction in an Additive Functional Model}
      \field{urlday}{06}
      \field{urlmonth}{10}
      \field{urlyear}{2013}
      \field{year}{2013}
      \verb{eprint}
      \verb 1301.4954
      \endverb
    \endentry
    \entry{wang2011smoothing}{book}{}
      \name{labelname}{1}{}{%
        {{hash=a2af46eff4389524fe951572da657010}{Wang}{W\bibinitperiod}{Y.}{Y\bibinitperiod}{}{}{}{}}%
      }
      \name{author}{1}{}{%
        {{hash=a2af46eff4389524fe951572da657010}{Wang}{W\bibinitperiod}{Y.}{Y\bibinitperiod}{}{}{}{}}%
      }
      \list{publisher}{1}{%
        {{CRC} Press}%
      }
      \strng{namehash}{a2af46eff4389524fe951572da657010}
      \strng{fullhash}{a2af46eff4389524fe951572da657010}
      \field{sortinit}{W}
      \field{labelyear}{2011}
      \field{labeltitle}{Smoothing Splines: Methods and Applications}
      \field{title}{Smoothing Splines: Methods and Applications}
      \field{year}{2011}
    \endentry
    \entry{wang2011testing}{article}{}
      \name{labelname}{2}{}{%
        {{hash=a2af46eff4389524fe951572da657010}{Wang}{W\bibinitperiod}{Y.}{Y\bibinitperiod}{}{}{}{}}%
        {{hash=742f8103481c80bc5dcf8cf53a3fff91}{Chen}{C\bibinitperiod}{H.}{H\bibinitperiod}{}{}{}{}}%
      }
      \name{author}{2}{}{%
        {{hash=a2af46eff4389524fe951572da657010}{Wang}{W\bibinitperiod}{Y.}{Y\bibinitperiod}{}{}{}{}}%
        {{hash=742f8103481c80bc5dcf8cf53a3fff91}{Chen}{C\bibinitperiod}{H.}{H\bibinitperiod}{}{}{}{}}%
      }
      \strng{namehash}{54e65a87440b9b5ed9a5f736044a8ddb}
      \strng{fullhash}{54e65a87440b9b5ed9a5f736044a8ddb}
      \field{sortinit}{W}
      \field{labelyear}{2012}
      \field{labeltitle}{On testing an unspecified function through a linear mixed effects model with multiple variance components}
      \field{journaltitle}{Biometrics}
      \field{number}{4}
      \field{title}{On testing an unspecified function through a linear mixed effects model with multiple variance components}
      \field{volume}{68}
      \field{year}{2012}
      \field{pages}{1113\bibrangedash 1125}
      \verb{doi}
      \verb 10.1111/j.1541-0420.2012.01790.x
      \endverb
    \endentry
    \entry{wood2006low}{article}{}
      \name{labelname}{1}{}{%
        {{hash=eabb4845fa2091976b67811dee89cf1f}{Wood}{W\bibinitperiod}{S.\bibnamedelimi N.}{S\bibinitperiod\bibinitdelim N\bibinitperiod}{}{}{}{}}%
      }
      \name{author}{1}{}{%
        {{hash=eabb4845fa2091976b67811dee89cf1f}{Wood}{W\bibinitperiod}{S.\bibnamedelimi N.}{S\bibinitperiod\bibinitdelim N\bibinitperiod}{}{}{}{}}%
      }
      \list{publisher}{1}{%
        {Wiley Online Library}%
      }
      \strng{namehash}{eabb4845fa2091976b67811dee89cf1f}
      \strng{fullhash}{eabb4845fa2091976b67811dee89cf1f}
      \field{sortinit}{W}
      \field{labelyear}{2006}
      \field{labeltitle}{Low-Rank Scale-Invariant Tensor Product Smooths for Generalized Additive Mixed Models}
      \field{journaltitle}{Biometrics}
      \field{number}{4}
      \field{title}{Low-Rank Scale-Invariant Tensor Product Smooths for Generalized Additive Mixed Models}
      \field{volume}{62}
      \field{year}{2006}
      \field{pages}{1025\bibrangedash 1036}
    \endentry
    \entry{wood2011fast}{article}{}
      \name{labelname}{1}{}{%
        {{hash=eabb4845fa2091976b67811dee89cf1f}{Wood}{W\bibinitperiod}{S.\bibnamedelimi N.}{S\bibinitperiod\bibinitdelim N\bibinitperiod}{}{}{}{}}%
      }
      \name{author}{1}{}{%
        {{hash=eabb4845fa2091976b67811dee89cf1f}{Wood}{W\bibinitperiod}{S.\bibnamedelimi N.}{S\bibinitperiod\bibinitdelim N\bibinitperiod}{}{}{}{}}%
      }
      \list{publisher}{1}{%
        {Royal Statistical Society}%
      }
      \strng{namehash}{eabb4845fa2091976b67811dee89cf1f}
      \strng{fullhash}{eabb4845fa2091976b67811dee89cf1f}
      \field{sortinit}{W}
      \field{labelyear}{2011}
      \field{labeltitle}{Fast stable restricted maximum likelihood and marginal likelihood estimation of semiparametric generalized linear models}
      \field{journaltitle}{Journal of the Royal Statistical Society: Series B (Statistical Methodology)}
      \field{number}{1}
      \field{title}{Fast stable restricted maximum likelihood and marginal likelihood estimation of semiparametric generalized linear models}
      \field{volume}{73}
      \field{year}{2011}
      \field{pages}{3\bibrangedash 36}
    \endentry
    \entry{wood2013p}{article}{}
      \name{labelname}{1}{}{%
        {{hash=eabb4845fa2091976b67811dee89cf1f}{Wood}{W\bibinitperiod}{S.\bibnamedelimi N.}{S\bibinitperiod\bibinitdelim N\bibinitperiod}{}{}{}{}}%
      }
      \name{author}{1}{}{%
        {{hash=eabb4845fa2091976b67811dee89cf1f}{Wood}{W\bibinitperiod}{S.\bibnamedelimi N.}{S\bibinitperiod\bibinitdelim N\bibinitperiod}{}{}{}{}}%
      }
      \list{publisher}{1}{%
        {Biometrika Trust}%
      }
      \strng{namehash}{eabb4845fa2091976b67811dee89cf1f}
      \strng{fullhash}{eabb4845fa2091976b67811dee89cf1f}
      \field{sortinit}{W}
      \field{labelyear}{2013}
      \field{labeltitle}{On p-values for smooth components of an extended generalized additive model}
      \field{journaltitle}{Biometrika}
      \field{number}{1}
      \field{title}{On p-values for smooth components of an extended generalized additive model}
      \field{volume}{100}
      \field{year}{2013}
      \field{pages}{221\bibrangedash 228}
    \endentry
    \entry{wood2012straightforward}{article}{}
      \name{labelname}{3}{}{%
        {{hash=eabb4845fa2091976b67811dee89cf1f}{Wood}{W\bibinitperiod}{S.\bibnamedelimi N.}{S\bibinitperiod\bibinitdelim N\bibinitperiod}{}{}{}{}}%
        {{hash=4f1e5e96ac9e3d1124e4f0be96e8c9f6}{Scheipl}{S\bibinitperiod}{F.}{F\bibinitperiod}{}{}{}{}}%
        {{hash=2c553fdb976ea52c6856dd7f2ef64d0f}{Faraway}{F\bibinitperiod}{J.\bibnamedelimi J.}{J\bibinitperiod\bibinitdelim J\bibinitperiod}{}{}{}{}}%
      }
      \name{author}{3}{}{%
        {{hash=eabb4845fa2091976b67811dee89cf1f}{Wood}{W\bibinitperiod}{S.\bibnamedelimi N.}{S\bibinitperiod\bibinitdelim N\bibinitperiod}{}{}{}{}}%
        {{hash=4f1e5e96ac9e3d1124e4f0be96e8c9f6}{Scheipl}{S\bibinitperiod}{F.}{F\bibinitperiod}{}{}{}{}}%
        {{hash=2c553fdb976ea52c6856dd7f2ef64d0f}{Faraway}{F\bibinitperiod}{J.\bibnamedelimi J.}{J\bibinitperiod\bibinitdelim J\bibinitperiod}{}{}{}{}}%
      }
      \list{publisher}{1}{%
        {Springer}%
      }
      \strng{namehash}{ecdb32663c76a71284ac2e71ef99e753}
      \strng{fullhash}{ddf724f7d4e149208c8d66fd49da10b9}
      \field{sortinit}{W}
      \field{labelyear}{2013}
      \field{labeltitle}{Straightforward intermediate rank tensor product smoothing in mixed models}
      \field{journaltitle}{Statistics and Computing}
      \field{number}{3}
      \field{title}{Straightforward intermediate rank tensor product smoothing in mixed models}
      \field{volume}{23}
      \field{year}{2013}
      \field{pages}{341\bibrangedash 360}
    \endentry
    \entry{zhang2007statistical}{article}{}
      \name{labelname}{2}{}{%
        {{hash=31f9cc14d335e93631c5ac98b3bf57e3}{Zhang}{Z\bibinitperiod}{J.-T.}{J\bibinithyphendelim T\bibinitperiod}{}{}{}{}}%
        {{hash=9be4d62c296f903e9a24a63cbbdb61c6}{Chen}{C\bibinitperiod}{J.}{J\bibinitperiod}{}{}{}{}}%
      }
      \name{author}{2}{}{%
        {{hash=31f9cc14d335e93631c5ac98b3bf57e3}{Zhang}{Z\bibinitperiod}{J.-T.}{J\bibinithyphendelim T\bibinitperiod}{}{}{}{}}%
        {{hash=9be4d62c296f903e9a24a63cbbdb61c6}{Chen}{C\bibinitperiod}{J.}{J\bibinitperiod}{}{}{}{}}%
      }
      \list{publisher}{1}{%
        {Institute of Mathematical Statistics}%
      }
      \strng{namehash}{a107b821c5f2801aabee587ad2ea08e8}
      \strng{fullhash}{a107b821c5f2801aabee587ad2ea08e8}
      \field{sortinit}{Z}
      \field{labelyear}{2007}
      \field{labeltitle}{Statistical inferences for functional data}
      \field{journaltitle}{The Annals of Statistics}
      \field{number}{3}
      \field{title}{Statistical inferences for functional data}
      \field{volume}{35}
      \field{year}{2007}
      \field{pages}{1052\bibrangedash 1079}
    \endentry
  \endsortlist
\endrefsection

  \blx@bblend
  \endgroup
  \csnumgdef{blx@labelnumber@\the\c@refsection}{0}}
\makeatother
\begin{document}

\title{
Restricted Likelihood Ratio Tests for Linearity in Scalar-on-Function Regression
}
%
\author{Mathew W. McLean\thanks{Research Assistant Professor, Institute for Applied Mathematics and Computational Science, Texas A\&M University, College Station, TX, 77843  (E-mail: mmclean@stat.tamu.edu)}
\and
Giles Hooker\thanks{Associate Professor, Department of Biological Statistics and Computational Biology, Cornell University, Ithaca, NY, 14853, USA (E-mail: giles.hooker@cornell.edu)}
\and
David Ruppert
\thanks{Andrew Schultz Jr.\ Professor of Engineering and
Professor of Statistical Science, School of Operations Research
and Information Engineering and Department of Statistical Science, Cornell University, 1170 Comstock Hall, Ithaca, NY
14853, USA (E-mail: dr24@cornell.edu) }
}

\date{\today}

\maketitle
\begin{abstract}
\noindent
 We propose a procedure for testing the linearity of a scalar-on-function regression relationship.  To do so, we use the functional generalized additive model (FGAM), a recently developed extension of the functional linear model.  For a functional covariate $X(t)$, the FGAM models the mean response as the integral with respect to $t$ of $F\{X(t),t\}$ where $F(\cdot,\cdot)$ is an unknown bivariate function.  The FGAM can be viewed as the natural functional extension of generalized additive models.  We show how the functional linear model can be represented as a simple mixed model nested within the FGAM.  Using this representation, we then consider restricted likelihood ratio tests for zero variance components in mixed models to test the null hypothesis that the functional linear model holds.  The methods are general and can also be applied to testing for interactions in a multivariate additive model or for testing for no effect in the functional linear model.  The performance of the proposed tests is assessed on simulated data and in an application to measuring diesel truck emissions, where strong evidence of nonlinearities in the relationship between the functional predictor and the response are found.
\end{abstract}
\section{Introduction}\label{LinearityTests}
The purpose of this paper is to propose and compare formal tests for when a scalar-on-function regression is well-modelled by the functional linear model (FLM).  The FLM is by far the most popular model for scalar-on-function regression with a number of applied and theoretical developments available in the literature; an overview is provided by \textcite{ramsay2005Functional}.  We suppose one has data $\{(X_i(t),Y_i):  \ t \in \mcT\}$ for $i=1,\ldots,N$, where $X_i$ is a real-valued, continuous, square-integrable, random function on the compact interval $\mcT$ and $Y_i$ is a scalar.  We make the common assumption that the predictor, $X(\cdot)$, is observed at a dense grid of points.  The FLM assumes the mean effect of $X$ on $Y$ is linear for each fixed $t$ as follows
\begin{equation}
Y_i=\beta_0+\int_{\mathcal{T}}\beta(t)X_i(t)\,dt+\eps_i,\label{flm}
\end{equation}
where $\beta(\cdot)$ is the functional coefficient with $\beta(t)$ describing the effect on the response of the functional predictor at time $t$ and $\eps_i\simiid N(0,\sigma^2)$.  

\textcite{McLean2012functional} proposed an extension of the FLM called the functional generalized additive model (FGAM), which assumes the following form for the response
\begin{align}
Y_i=\beta_0+\int_\mcT F\{X_i(t),t\}\,dt+\eps_i,\label{fgam}
\end{align}
where $F(\cdot,\cdot)$ is an unknown bivariate function that is assumed to be smooth.  The FLM can be seen as a special case of FGAM where $F(x,t)=\beta(t)x$.  An intuitive way to view~\eqref{fgam}, is to consider it as the limit as the number of predictors (i.e.\ the number of observation times, $t_j$) goes to infinity in a (multivariate) additive model with component functions $f_j\{X_i(t_j)\}\Delta t_j:=F\{X_i(t_j),t_j\}$.  Each additive component can be seen as a term in the Riemann sum approximation to the integral in~\eqref{fgam} \autocite[see,][]{McLean2012functional}.  Some asymptotic theory for different estimators of \eqref{fgam} is available in \textcite{wang2013optimal,mueller2013continuously}.  An extension to the case of sparsely observed functional predictors is also available \autocite{mclean2013bayesian}.

We briefly review some other nonlinear scalar-on-function regression models that have been proposed in the literature.  One of the earliest such proposals, and perhaps the most general, is the nonparametric kernel estimator of \textcite{ferraty2006Nonparametric}.  Several authors have considered additive models that use linear functionals of the predictor curves as covariates, e.g.\
\[
E(Y_i\mid X_i)=\beta_0+f\{\langle\beta(t)X_i(t)\rangle\}=\beta_0+f\{\int\beta(t)X_i(t)dt\},
\]
 for unknown $\beta_0,\ f(\cdot)$, and $\beta(t)$.  Two such examples are \textcite{muller2008functional} and \textcite{james2005functional}. The former approach regresses on a finite number of functional principal components scores and the latter approach searches for linear functionals using projection pursuit.  For ease of interpretation, FGAM may be preferred to these models as it incorporates the functional predictor directly in the conditional mean; a model that is additive in the principal component scores is not additive in $X(t)$ itself, and vice versa.  While these models and FGAM each have the FLM as a special case, each is capable of capturing true relationships for the conditional mean response that the others cannot, and they can thus be seen as complimentary.  Single-index type models similar to \textcite{james2005functional} are considered in \textcite{ait2008cross,chen2011single,febrero2013gam}.  Two other noteworthy extensions of the FLM are \textcite{guillas2010bivariate}, which examines the case when $X$ is a bivariate function so that $E(Y_i\mid X_i)=\beta_0+\int\int\beta(s,t)X(s,t)\,dsdt$ and \textcite{li2010generalized}, which allows for interaction between a scalar and functional covariate though a single index. 

This paper will focus on testing
\begin{align*}
H_0&:\ E(Y_i\mid X_i)=\theta_0+\int_\mcT\beta(t)X_i(t)\,dt\quad (\text{FLM})\\
 \intertext{vs.}
H_1&:\ E(Y_i\mid X_i)=\theta_0+\int_\mcT F(X_i(t),t)\,dt\quad (\text{FGAM}).
\end{align*}
To accomplish this, we will formulate FGAM as a mixed model using a representation proposed for penalized splines in \textcite{wood2012straightforward}, obtaining a representation of FGAM as a mixed model with three pairwise independent vectors of random effects.  Mixed model representations for penalized splines are popular for their conceptual simplicity and because they can be estimated using software and algorithms available for mixed models \autocite[see, e.g.,][]{ruppert2003semiparametric}.  This parameterization makes the nesting of the FLM within FGAM explicit, and reduces our problem to one of testing for zero variance components in a mixed model.  To conduct the test for zero variance components, we use restricted likelihood ratio tests (RLRTs) for which the exact finite-sample distribution is known for the one variance component case \autocite{crainiceanu2004Likelihood}.

Though the papers mentioned above all propose nonlinear models for scalar-on-function regression, we are only aware of one other work \autocite{febrero2013goodness} that considers a formal test of the FLM being true under the null hypothesis.  To date almost all hypothesis tests proposed for scalar-on-function regression are concerned with testing for no effect in the FLM (i.e.\ $H_0:\ \beta\equiv 0$ in \eqref{flm}).  Examples include \textcite{cardot2003testing,swihart2013restricted,gabrys2010tests,staicu2013classical,zhang2007statistical}.  \textcite{swihart2013restricted} is notable for taking a similar approach to the one we consider here; they use a penalized spline-mixed model for the FLM and use an RLRT for a zero variance component to test their desired hypothesis.  Other hypothesis tests available for scalar-on-function regression include a test for equality of two coefficient functions estimated from independent data sets and a test for no quadratic effect in the functional quadratic model \autocite[Ch.~10 and 12, resp.]{horvath2012inference}.

\textcite{febrero2013goodness} proposes a Cram\'{e}r-von Mises statistic for testing the null hypothesis that the FLM is the true model and uses a wild bootstrap to approximate the null distribution of the statistic.  The method estimates the coefficient function in the FLM using a finite number of basis functions without penalization and has an unspecified alternative hypothesis.  In contrast, we use \textit{penalized} splines which allow for greater flexibility and specify an alternative hypothesis that the FGAM is the true model.  We will compare our RLRT with the test used by these authors in our numerical studies.

The remainder of the paper proceeds as follows.  In Section~\ref{sec_mmrep} we discuss mixed model representations for FGAM.  Section~\ref{sec_FGAMtest} presents our proposed restricted likelihood ratio tests for linearity of FGAM.  Section~\ref{sec_MMsimstudy} provides a simulation study of our proposed approaches, and Section~\ref{sec_Emissions} concludes with an application of our methods to some motor vehicle emissions data.
\section{Mixed Models Representations for FGAM}\label{sec_mmrep}
\subsection{Notation and Basic Mixed Model Parameterization}\label{sec_BasisCons}
In this section we introduce notation and briefly discuss a simple mixed model representation for FGAM in order to motivate the alternative mixed model representation that will be the focus of the paper.  The linear mixed model we present here uses the parameterization of the bivariate surface in \eqref{fgam} that was used by \textcite{McLean2012functional}, which we now describe.  For both the $x$ and $t$ axes, we define cubic B-spline bases which we denote by $\{B_j^X(x):\ j=1,\ldots,K_x\}$ and $\{B_k^T(t):\ k=1,\ldots,K_t\}$, respectively.  For simplicity, the knots for the bases are chosen to be equally-spaced distances apart.  The surface in \eqref{fgam} becomes
\[
F(x,t)=\sum_{j=1}^{K_x}\sum_{k=1}^{K_t}B_j^X(x)B_k^T(t)\theta_{jk},
\]
where the $\theta_{jk}$'s are unknown tensor-product B-spline coefficients.

Let $\matB_x$  denote the $NJ\times K_x$ matrix of the $x$-axis B-splines evaluated at vec$(\vecX)$, where $\vecX$ is the $N\times J$ matrix of observed functional predictor values ($N$ curves measured $J$ times each).  Similarly, define $\matB_t$ to be the $NJ\times K_t$ matrix of $t$-axis B-splines evaluated at $\text{vec}(\vecT)$, where  $\vecT$ is the $N\times J$ matrix of observation times for the functional predictor.  In the densely-observed functional predictor setting assumed for this work, each $X(t)$ will be pre-smoothed, and the statistician is free to choose $\vecT$ and $J$ as he/she chooses.  We use a second-order penalty for both the $x$ and $t$ axes, $\int\{\partial_{xx}F(x,t)\}^2\,dx$ and $\int\{\partial_{tt}F(x,t)\}^2\,dt$, respectively.   To form the tensor product surface, we use the box product, also known as the row-wise Kronecker product, which is defined for two matrices $\matA_1$ and $\matA_2$ of dimension $n\times m_1$ and $n\times m_2$, respectively as $\matA_1\Box\matA_2:=\matA_1\otimes\vecone^T_{m_1}\odot\vecone^T_{m_2}\otimes\matA_2$, where $\otimes$ represents the Kronecker product and $\odot$ denotes element-wise matrix multiplication.  We must approximate the integral in \eqref{fgam} using numerical quadrature.  For this we define an $N\times NJ$ matrix $\matL$ of quadrature weights. For example, if we simply use the midpoint rule, $\matL=J^{-1}(\matI_N\otimes\vecone^T_J)$.  The penalized log-likelihood for FGAM is then
 \beq\label{plik}
-\frac{N}{2}\log(\sigma^2)- \frac{1}{2\sigma^2}\lVert\vecy-\vecone\beta_0-\matL\matB_x\Box\matB_t\btheta\rVert^2-\btheta^T\matS(\lambda_x,\lambda_t)\btheta
 \eeq
 where $\vecy=(y_1\ldots,y_N)^T$, $\btheta=(\theta_{11},\ldots,\theta_{1K_t},\theta_{21},\ldots,\theta_{K_xK_t})^T$, and $\matS(\lambda_x,\lambda_t)$ is the penalty matrix for the tensor product smooth \autocite[e.g.,][]{wood2006low,marx2005multidimensional}. 

 \textcite{McLean2012functional} used generalized cross validation (GCV) to select the smoothing parameters $\lambda_x$ and $\lambda_t$, which is equivalent to placing an improper Gaussian prior on the spline coefficients.  To fit FGAM using mixed modelling software, we require a proper Gaussian prior for the random effects.  This can be achieved by separating the model into a component in the null space of the penalty term for FGAM and a component in the orthogonal complement of the null space, which we refer to as the range space for consistency with \textcite{wood2012straightforward}.  One easy way to do this is using an eigendecomposition of the penalty matrix.

 We decompose $\matS(\lambda_x,\lambda_t)$ as $\matS(\lambda_x,\lambda_t)=\matU\matD\matU^T$, where $\matU$ is orthogonal and $\matD$ is diagonal.   Owing to the second-order penalties, $\matS(\lambda_x,\lambda_t)$ will have four zero-eigenvalues.  We split $\matU$ into $\matU=[\matU_p:\matU_n]$, where $\matU_{n}$ contains the columns of $\matU$ corresponding to the zero eigenvalues of the penalty matrix and $\matU_{p}$ consists of the remaining columns.  We write $\matD_+$ to denote the diagonal submatrix of $\matD$ containing only the positive eigenvalues of $\matS(\lambda_x,\lambda_t)$.   The columns of $\matU_p$ form a basis for the range space of the FGAM penalty.  FGAM can then then be reparametrized as follows
  \[
\begin{pmatrix}\vecb\\\bbeta\end{pmatrix}:=\matU^T\btheta,\quad \matX:=\matL\matB_x\Box\matB_t\matU_{n},\quad \matZ:=\matL\matB_x\Box\matB_t\matU_{p},
  \]
  and we then have the mixed model
 \begin{gather} \label{altFGAMM}
 \vecy=\vecone\beta_0+\matX\bbeta+\matZ\vecb+\bepsilon,\\
 \bepsilon\sim N(\veczero,\sigma^2\matI_N),\quad \vecb\sim N\left(\veczero,[\matU_p^T\matS(\lambda_x,\lambda_t)\matU_p]^{-1}\right).\nonumber
 \end{gather}
 We can further reparametrize \eqref{altFGAMM} in terms of variance components $\sigma_x^2:=\sigma^2/\lambda_x$ and $\sigma_t^2=\sigma^2/\lambda_t$ to coincide with the mixed model literature.

 This model can be fit via maximum likelihood (ML) or restricted maximum likelihood (REML) using some mixed modelling software such as the \texttt{nlme} package in \texttt{R} \autocite{pinheiro2013nlme}, but not by other popular software such as \texttt{SAS}\autocite{procmixed2012} or \texttt{R} package \texttt{lme4} \autocite{bates2013lme4} because of the complicated covariance structure for the random effect.  There is some evidence to suggest that REML should be preferred to methods that minimize prediction error, such as GCV or Akaike's information criteria (AIC) and its siblings for estimating smoothing parameters in penalized spline models because it avoids their occasional tendencies to badly undersmooth and offers slightly better RMSE performance in practise \autocite{reiss2009smoothing}.  However, for the purposes of this work, we need a model with a simpler covariance structure that makes the nesting of the FLM within FGAM explicit.  This will be the subject of the next section.
\subsection{Penalized Spline ANOVA Parameterization}\label{sec_psanova}
In this section we show how FGAM may be parameterized using the tensor product spline basis construction of \textcite{wood2012straightforward} which parallels smoothing spline ANOVA \autocite[for e.g.,][]{wang2011smoothing}, the main difference being the use of low-rank spline bases.  The full model will be projected onto a tensor sum of orthogonal subspaces, with each component in the new construction either unpenalized or with its own unique penalty that is interpretable in terms of the original model component functions.  This is convenient because it will lead to a mixed model representation where each random effect has a diagonal covariance matrix independent of the other effects.

For this construction, we begin with eigendecompositions of marginal penalty matrices, $\matP_x$ and $\matP_t$.  For a second-order penalty, $\matP_x$ has rank $K_x-2$ and $(n,m)$-entry
\[
(P_x)_{m,n}=\int \partial_{xx}\{B^{(X)}_m(x)\}\partial_{xx}\{B^{(X)}_n(x)\}\,dx;\ m,n=1,\ldots,K_x.
\]
We express $\matP_x$ as $\matP_x=\matU_x\matD_x\matU_x^T$, where $\matU_x$ is orthogonal and $\matD_x$ is diagonal with two zeros on the diagonal.  As before, we let $\matU_{n,x}$ be the columns of $\matU_x$ corresponding to the zero eigenvalues and $\matU_{p,x}$ be the remaining columns and let $\matD_{+,x}$ be the diagonal matrix containing all positive eigenvalues of $\matP_x$.  We then form $\matZ_x=\matB_x\matU_{p,x}\matD^{-1/2}_{x,+}$ and $\matX_x=\matB_x\matU_{n,x}$.  The matrix $\matZ_x$ forms a basis for the random effects of a univariate smooth in $x$ (i.e., a basis for the range space of the marginal penalty for $x$) and $\matX_x$ forms a basis for the fixed effects of the smooth.  The marginal penalty matrix will become the identity matrix of appropriate dimension except with its last two diagonal entries equal to zero \autocite[see][]{wood2012straightforward}.  We form $\matX_t$ and $\matZ_t$ analogously using $\matB_t$ and a corresponding penalty matrix $\matP_t$.

To obtain our tensor product construction, we form all pairwise box products of elements of $\{\matX_x,\ \matZ_x\}$ with elements of $\{\matX_t,\ \matZ_t\}$.  Our design matrix for the tensor product smooth becomes
\[
\mathcal{M}=\left[\matX_x\Box\matX_t:\matX_x\Box\matZ_t:\matZ_x\Box\matX_t:\matZ_x\Box\matZ_t \right].
\]
The term $\matX_x\Box\matX_t$ corresponds to the unpenalized, fixed effects part of the smooth, and the three other terms are bases for the random effects with each component having a separate ridge penalty.  Let $\vecx=\text{vec}(\vecX)$ and $\vect=\text{vec}(\vecT)$; we can re-parameterize the null space bases as $\matX_x=[\vecone:\vecx]$, $\matX_t=[\vecone:\vect]$, and $\matX_x\Box\matX_t=[\vecone:\vecx:\vect:\vecx\odot\vect]$.  The function $F(x,t)$ is decomposed into an unpenalized, parametric part and three separate nonparametric parts each subject to different, non-overlapping penalties
\begin{align}\label{FGAMdecomp}
&\underbrace{\text{term}}_{\text{penalty}}:\underbrace{F(x,t)}_{\lambda_t\int(\partial_{tt}F)^2+\lambda_x\int(\partial_{xx}F)^2}=\underbrace{\beta_0+\beta_1x+\beta_2t+\beta_3xt}_{\text{unpenalized}}\nonumber\\
&+\underbrace{f_1(t)+xf_2(t)}_{\lambda_1[\int(\partial_{tt}f_1)^2+(\partial_{tt}f_2)^2]}
\qquad+ \underbrace{g_1(x)+tg_2(x)}_{\lambda_2[\int(\partial_{xx}g_1)^2+(\partial_{xx}g_2)^2]}
+ \underbrace{h(x,t)}_{\lambda_3\int(\partial_{xxtt}h)^2}.
\end{align}

\begin{table}
\centering
\caption[Description of tensor product construction for linearity tests]{Description of penalized components of the tensor production construction (\ref{FGAMdecomp}) with $f_1(t)$ removed for identifiability\label{pencomp}}
\begin{tabular}{|l|c|c|}
\hline\noalign{\smallskip}
& Basis for functions & \\
Term & of the form & Penalty \\
\noalign{\smallskip}\hline\noalign{\smallskip}
$\matX=[\vecone:\vecx:\vecx\otimes\vect]$ & $\beta_0+\beta_1x+\beta_3xt$  & unpenalized \\
\noalign{\smallskip}\hline\noalign{\smallskip}
$\matZ_1=\vecx\Box\matZ_t$ & $xf_2(t)$  & $\int(\partial_{tt}f_1)^2+(\partial_{tt}f_2)^2$ \\
\noalign{\smallskip}\hline\noalign{\smallskip}
$\matZ_2=\matZ_x\Box\matX_t$ & $g_1(x)+tg_2(x)$ &  $\int(\partial_{xx}g_1)^2+(\partial_{xx}g_2)^2$ \\
\noalign{\smallskip}\hline\noalign{\smallskip}
$\matZ_3=\matZ_x\Box\matZ_t$ & $h(x,t)$ w/o above terms &  $\int(\partial_{xxtt}h)^2$ \\
\noalign{\smallskip}\hline
\end{tabular}
\end{table}
 If the terms $\beta_2t$ and $f_1(t)$ in \eqref{FGAMdecomp} are integrated w.r.t.\ $t$, they become confounded with the intercept, $\beta_0$.  Hence, for identifiability $\vect$ and $\matZ_t\ (=\vecone\Box\matZ_t)$ are dropped from $\mcM$.  This construction is summarized in Table~\ref{pencomp}.  An alternative approach for ensuring identifiability is discussed in \textcite{McLean2012functional}.  If we define
 \[
 \matX=[\vecone:\vecx:\vecx\otimes\vect],\ \matZ_1=\vecx\Box\matZ_t,\ \matZ_2=\matZ_x\Box\matX_t,\ \matZ_3=\matZ_x\Box\matZ_t,
  \]
  and let $\mathcal{L}(\vecF)$ denote the $N$-vector with $i$th entry\\ $\int_\mcT F\{X_i(t),t\}\,dt$, then we can write FGAM in mixed model form as
\begin{align}
\vecy&=\beta_0+\mathcal{L}(\vecF)\approx\matL\matX\bbeta+\sum_{j=1}^3\matL\matZ_j\vecb_j+\bepsilon;\label{lmeform}\\
\vecb_j&\sim N(\veczero,\sigma_{j}^2\matI_{q_j}),\ j=1,2,3;\notag\\
\bepsilon&\sim N(\veczero,\sigma^2_e\matI_N),\notag
\end{align}
with the dimension of the vectors of random effects being $q_1=K_t-2,\ q_2=2(K_x-2),\text{ and }q_3=(K_x-2)(K_t-2)$ for $\vecb_1,\ \vecb_2,\text{ and }\vecb_3$, respectively.

We now have a mixed model where each vector of random effects, $\vecb_j$, has a simple, diagonal covariance structure.   This allows us to use \texttt{lme4}, which is optimized to work with sparse covariance matrices, or e.g., \texttt{PROC MIXED} in \texttt{SAS}.  This is not possible for the mixed model \eqref{altFGAMM}.  More importantly for the goal of this paper, it is explicitly clear how the FLM is nested in FGAM in this parameterization: Referencing Table~\ref{pencomp}, the integrand for the FLM, $\beta(t)X(t)$, can be seen to correspond to the first penalized term, $xf_2(t)$, and the unpenalized terms.  Thus, in the mixed model \eqref{lmeform}, the FLM is represented by the fixed effects and the random effect $\vecb_1$.  Therefore, a goodness of fit test for the FLM is equivalent to testing whether $\sigma_2=\sigma_3=0$.  We will refer to \eqref{lmeform} as the FGAMM (for Functional Generalized Additive Mixed Model) later in our numerical experiments when we compare this parametrization with the one used in \textcite{McLean2012functional}.
\section{Tests for Linearity of Scalar-on-Function Regression}\label{sec_FGAMtest}
\subsection{Approximate Restricted Likelihood Ratio Tests}\label{sec_reml}
The restricted likelihood \autocite{patterson1971recovery} for model \eqref{lmeform} is
 \begin{align*}
 \ell_R(\vecy)&= -\frac{N}{2}\log(2\pi)-\frac{1}{2}\log\lvert\bSigma_J\rvert\\
 &-\frac{(\vecY-\matL\matX\bbeta)^T\bSigma^{-1}_J(\vecY-\matL\matX\bbeta)}{2\sigma^2}
 -\frac{1}{2}\lvert\matX^T\matL^T\bSigma_J^{-1}\matL\matX\rvert,
 \end{align*}
 where $\bSigma_J:=\text{cov}(\vecY)=\sigma^2\matI_N+\sum_{j=1}^3\sigma^2_j\matL\matZ_j\matZ_j^T\matL^T$.  We wish to test $H_0:\ \sigma_2=\sigma_3=0$ using a restricted likelihood ratio test (RLRT).  For the one variance component linear model $\vecy=\matX\beta+\matZ\vecb$ with $\vecb\sim N(\veczero,\sigma_1^2\matI_{q_1})$, we know the \textit{exact} finite-sample distribution for the RLRT statistic \autocite{crainiceanu2004Likelihood}:
\begin{align}\label{rlrt}
&\text{RLRT}=2\sup_{H_1}\ell_R(\vecy)-2\sup_{H_0}\ell_R(\vecy)\nonumber\\
&=\sup_{\lambda}\left[(N-q_0-1)\log\{1+U_n(\lambda)\}-\sum_{k=1}^{q_1}\log(1+\lambda\mu_{k})\right];
\end{align}
where $\lambda=\sigma_1^2/\sigma^2$ and $U(\lambda)=N(\lambda)/D(\lambda)$ with
\begin{align*}
N(\lambda)&=\sum_{k=1}^{q_1}\frac{\lambda\mu_k}{1+\lambda\mu_k}w_k^2,\\
D(\lambda)&=\sum_{k=1}^{q_1}\log(1+\lambda\mu_k)+\sum_{k=q_1+1}^{N-q_0}w_k^2;
\end{align*}
for $w_k\simiid N(0,1);\ k=1,\ldots,N-q_0-1$.  The $\mu_k$ are the eigenvalues of the matrix $\matZ^T(\matI_N-\matX(\matX^T\matX)\matX^T)\matZ$.  This distribution may be simulated from very quickly.  The eigendecompostion of the $q_1\times q_1$ matrix need only be computed once, and then all that is required to obtain a draw from the RLRT distribution is simulation of $q_1$ $\chi^2_1$ random variables and one $\chi^2_{N-q_0-q_1-1}$ random variable.

 While the theory is fully developed for the one variance component case, extensions to tests for models with multiple variance components (which we will require for FGAM) have proven much harder, and this is still an open problem.  An approach that has proven to work well empirically is that of \textcite{greven2008Restricted}, which used ideas from pseudo-likelihood estimation and relied on the assumption that the restricted likelihood ratio tests (RLRT) for their variance components of interest could be accurately approximated by an RLRT that assumes the nuisance random effects are known.  Another possible approach has been proposed by \textcite{wang2011testing}, which developed F-tests for penalized spline models estimated in the mixed model framework.  The Wald-type test developed by \textcite{wood2013p} is not considered because it was designed for testing whether a component of a smooth is identically zero, not for testing whether the component is in the null space of its penalty.  We choose to work with the approach of \textcite{greven2008Restricted} because it has been shown through extensive simulation studies by \textcite{scheipl2008Size} to work well and because the method is available in an \texttt{R} package by the same authors.  The simulations in \textcite{wang2011testing} also confirmed the effectiveness of the \textcite{greven2008Restricted} approach, with their F-tests only offering minor improvements in the case where a nuisance variance component is very close to zero.  We refer to the \textcite{greven2008Restricted} method as the pseudo-RLRT.  A more computationally intensive approach, would be to approximate the null distribution of the RLRT statistic using a parametric bootstrap as in \textcite{pinheiro2000linear}.

 An additional open question not addressed by the above papers is how to test for multiple variance components being simultaneously zero under the null hypothesis.  This is the situation that we are faced with for FGAM and the subsequent sections will propose some ideas for how to deal with this problem.  
\subsubsection{Test $\sigma_2$ and $\sigma_3$ Separately}\label{Bonferroni}
The first approximation we consider is to conduct two separate RLRTs for the two non-FLM variance components.  In the first test, the random effect $\vecb_3$ is fixed as the zero vector under both the null and alternative hypotheses, $H_0:\ \sigma_2=0,\sigma_3=0$ and $H_1:\ \sigma_2>0,\sigma_3=0$.  In the second test, $\vecb_2\equiv\veczero$ and $\sigma_3$ is tested for equality with zero.  Both tests still involve one nuisance variance component and so we use the approximate RLRT from \textcite{greven2008Restricted} for each test and then apply a Bonferroni correction to account for multiple testing.  We refer to the approach as method ``Bonferroni'' in later sections.
\subsubsection{Test assuming $\sigma_2=\sigma_3$}
\label{twoVC}
The second approximation we consider is to assume that the amount of smoothness for each non-FLM component of the decomposition in \eqref{FGAMdecomp} are equal, i.e. that $\sigma_2=\sigma_3$. Using this assumption, we are again reduced to testing one variance component for equality to zero in the presence of one nuisance variance component, for which we use the method of \textcite{greven2008Restricted}.  However, unlike the previous section, we only need to conduct one test.  We
refer to the approach as method ``EqualVC'' in later sections.
\subsection{A Test For No Effect In the Functional Linear Model} \label{sec_FLMtest}
Before assessing whether an FLM or FGAM provides a better fit to the data, one will want to determine whether the functional predictor has any effect on the response at all.  This is quite simple to test in our framework.  By simply dropping the random effects $\vecb_2$ and $\vecb_3$, we can test for no effect by considering $H_0:\ \beta_2=\beta_3=0,\ \sigma_1=0$ versus $H_1:\beta_2\neq 0\text{ or }\beta_3\neq 0\text{ or }\sigma_1>0$ (FLM is true).  The exact distribution of the LRT statistic for this test is known due to \textcite{crainiceanu2004Likelihood}.  Note that a restricted likelihood ratio test is inappropriate here because the fixed effects are different under the two hypotheses.  One can also use either an LRT or RLRT to test $H_0:\ \sigma_1=0$ vs. $H_1:\ \sigma_1>0$ which is a test that the effect of $X(t)$ is linear in $t$; $Y_i=\beta_0+\vecL^T\{\vecx_i\odot(\beta_1+\beta_2\vect)\}$.  If one instead uses a first order penalty for $x$ and $t$, then a test for no effect is equivalent to testing $\sigma_1=0$. This proposal is similar to one recently considered in \textcite{swihart2013restricted} for the penalized functional regression model of \textcite{goldsmith2011penalized}.  Those authors first perform a functional principal components analysis to estimate the predictor trajectories and then estimate the coefficient function in the FLM using penalized splines with a first-order difference penalty and different mixed model representation than the one considered here. It is also possible to test for a quadratic effect of the form $\int\zeta(t)X^2(t)dt$ if one uses a third order penalty for the marginal basis for $x$.
\section{Simulation Study}\label{sec_MMsimstudy}
In this section we study the performance of our proposed tests for linearity of FGAM on simulated data for two different setups.  First, in Section~\ref{sec_simCC}, we generate the response variable using a convex combination of an FLM and an FGAM in the functional predictor.  This is done to assess the size of the departure from linearity that our tests can detect in a way that is interpretable in terms of the original models.  In Section~\ref{sec_simMM}, we assess empirical Type I error rates and power for our tests by generating the response from the mixed model in Section~\ref{sec_psanova} for several different values of the variance components and compare performance with tests that know the value of the nuisance parameters.

We also consider the recently proposed method of \textcite{febrero2013goodness} mentioned in the introduction.  This is the only work besides ours we our aware of that focuses on a null hypothesis of the FLM being true.  Their method is implemented in the \texttt{R} package \texttt{fda.usc} \autocite{febrero2013package}.  To perform their test, their software first fits an FLM.  In our simulation studies, specifying more than four basis functions for the functional coefficient produced singularities in the model matrix that caused their software to fail.  We speculate that this is due to the lack of regularization in the method.  We therefore only report results for the four basis function case for this method, which we label ``GPGMFB''.

 For the methods that involve RLRTs, computations are done in \texttt{R} \autocite{r2012} using the package \texttt{RLRsim} \autocite{scheipl2008Size}.  The package requires fitted model objects for the model under both hypotheses, as well as a fit to the data with nuisance variance components equal to zero.  These fits are obtained using the package \texttt{lme4} \autocite{bates2013lme4}.
\subsection{True Model as Convex Combination of FLM and FGAM}\label{sec_simCC}
\begin{figure}
\centering
\includegraphics[width=.7\columnwidth]{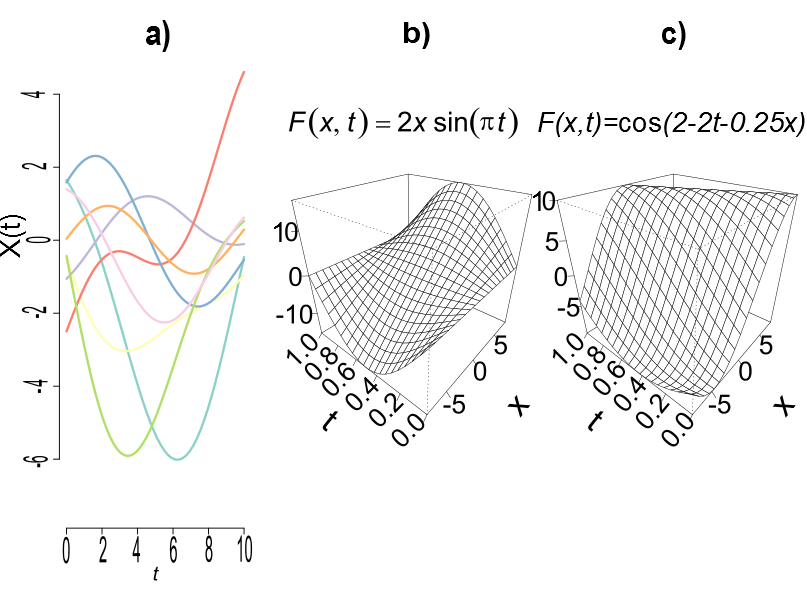}
\caption[Simulated Data]{\small a) Sample of 8 simulated predictor curves, $X(t)$.  b) True surface, $F_1(x,t)$, corresponding to the FLM.  c) True surface $F_2(x,t)$, corresponding to FGAM\label{simdata}}
\end{figure}

Here, we fit each model to 500 simulated data sets.  The true functional covariates are given by $X(t)=\sum_{j=1}^{4}\xi_j\phi_{j}(t),$ with $\xi_j\sim N(0,8j^{-2})$ and $\{\phi_1(t),\ldots,\phi_4(t)\}=\{\sin(\pi t),\cos(\pi t),$ $\sin(2\pi t),\cos(2\pi t)\}$.  Each functional predictor was observed at 30 equally-space points.  To generate the response, we take a convex combination of a bivariate function linear in $x$ and one nonlinear in $x$,
\[
F_1(x,t)=2x\sin(\pi t),\text{ and } F_2(x,t)=10\cos\left(-\frac{x}{8}+\frac{t}{4}-5\right),
\]
with $t=[0,1]$.  The response is given by
\[
Y_i=\int_0^1\left[\phi F_1\{X_i(t),t\}+(1-\phi)F_2\{X_i(t),t\}\right]dt+\epsilon_i,
\]
with $\epsilon_i\sim N(0,1)$ and $0\leq\phi\leq 1$.   The constants in $F_1$ and $F_2$ were chosen so that each surface contributed roughly equally to the signal for each generated data set prior to multiplication by $\phi$.  Notice that when $\phi=1$, the null hypothesis that the true model is an FLM is true.  Both true surfaces along with some generated functional predictors are shown in Figure~\ref{simdata}.

\begin{figure}
\centering
\includegraphics[width=.7\columnwidth]{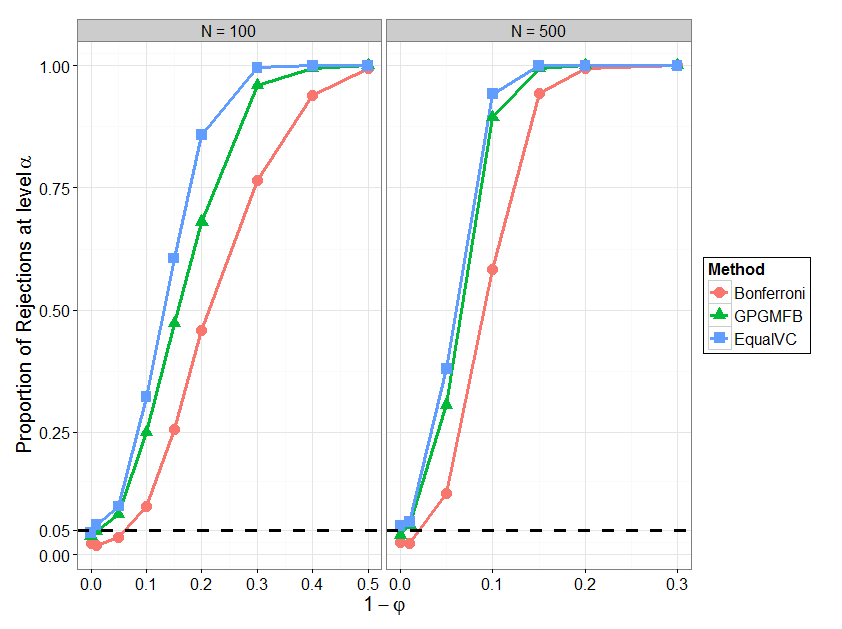}
\caption[Results for Section~\ref{sec_simCC} simulation study of testing methods]{\small Proportion of rejected null hypotheses for each testing method under consideration in Section~\ref{sec_simCC}\label{simCCres}}
\end{figure}

We consider two sample sizes, $N=100,500$.  We used ten basis functions for each axis when fitting the FGAM.  Results for other numbers of basis functions were similar, unless the number of basis functions is made too small.  This is because low-dimensional random effects are difficult for mixed model software to estimate.  In \texttt{lme4}, having low-dimensional random effects results in an increased number of zero estimates for the variance components corresponding to the low-dimensional random effects.  The results for this simulation study are summarized in Figure~\ref{simCCres} where for each of our proposed tests we plot the proportion of the 500 simulations where the null hypothesis is rejected by $(1-\phi)$.  We use $1-\phi$ so that zero on the x-axis corresponds to the null model (FLM) being true.  For brevity, we report results for significance level, $\alpha=0.05$ only.

The EqualVC method is able to achieve a type I error rate fairly close to the nominal level and also has the highest power of any of the methods.  Method Bonferroni is seen to be conservative, as expected.  GPGMFB has lower power and an observed type I error rate slightly further from the nominal level than method EqualVC, but is less conservative than Bonferroni.
\subsection{True Model as Mixed Model From Section~\ref{sec_psanova}}\label{sec_simMM}
We now change how the response is generated so that it comes from the mixed model~\eqref{lmeform}.  We will explore the trade-off between $\sigma_2^2$ and $\sigma_3^2$ values to assess power and to explore whether the assumption of EqualVC that $\sigma_2=\sigma_3$ can be problematic.  The functional predictors are generated in the same manner as the previous section.  For a given simulated sample of $N$ curves, the response is formed as follows.  First, the Section~\ref{sec_psanova} parameterization is used to form the bases $\matX,\ \matZ_1,\ \matZ_2,\text{ and }\matZ_3$.  Next, the random effect vector for the nuisance variance component corresponding to the FLM term (see Table~\ref{pencomp}) in the construction is drawn as $\vecb_1\sim N(\veczero, 4\matI_{q_1})$ and the two random effects vectors corresponding to non-FLM terms are drawn as $\vecb_j\sim N(\veczero, \sigma_j^2\matI_{q_j});\ j=2,3$.  The response is than given by
\[
\vecY=\matX\bbeta+\sum_{j=1}^3\matZ\vecb_j+\bepsilon,
\]
where $\bepsilon\sim N(\veczero,\matI_{N})$ and $\bbeta=(1,0.01,0.01)^T$.

In this section we again consider Bonferroni and EqualVC, and to better assess the performance of method Bonferroni, we consider a ``quasi-oracle'' test that knows the true value of the nuisance random effects vector for each simulation.  In more detail, the pseudo-residuals used as inputs to the pseudo-RLRTs for this method are $\vecY-\matZ_1\vecb_1$, for the true $\vecb_1$ instead of its prediction using REML.  The method still tests $\sigma_2$ and $\sigma_3$ separately and uses a Bonferroni correction, but with no nuisance variance component and only one variance component to test at a time, we are in exactly the framework of \textcite{crainiceanu2004Likelihood}, where the distribution of the test statistic is known and easily simulated from.  We label this method ``KnownSig1''.  Note also that changing the number of basis functions with this data generation scheme changes the dimension of the random effects in the true model.  The values of $\sigma_j^2$ for $j=2,3$; considered are $\sigma_j^2=(0,0.04, 0.1, 0.25, 0.5, 0.75)$ for $N=100$ and $\sigma_j^2=(0, 0.004, 0.04, 0.14, 0.2, 0.3)$ when $N=500$.

\begin{figure}[h]
\centering
\includegraphics[width=.85\columnwidth]{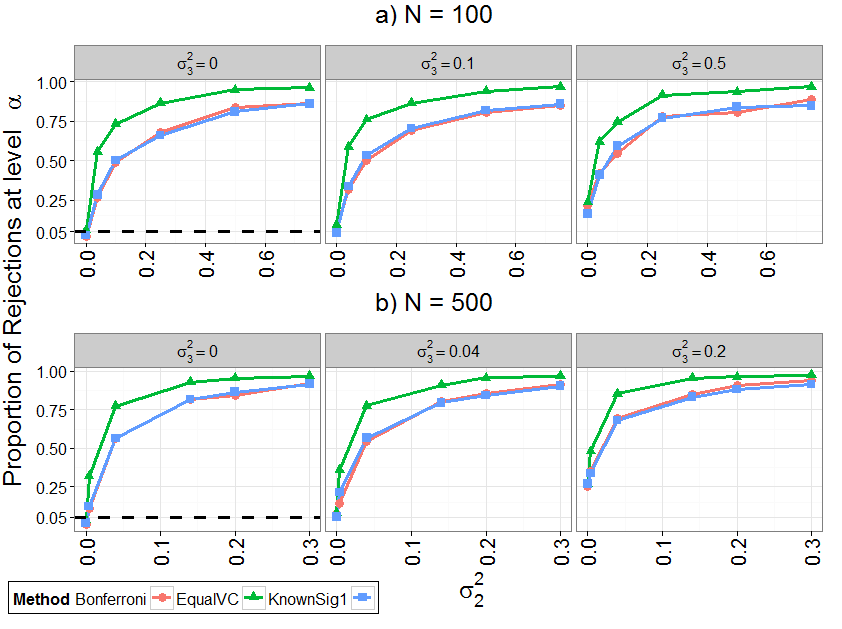}
\caption[Power of linearity tests for Section~\ref{sec_simMM} simulation study]{\small Proportion of rejected null hypotheses at $\alpha=0.05$ over 500 simulations for a grid of several values of $\sigma_2^2$ and $\sigma_3^2$\label{simMMres}}
\end{figure}

As in the previous section, we generate 500 data sets for each simulation setting, use ten basis functions for each axis when fitting FGAM, and report the proportion of times each method rejects the null hypothesis that the FLM is the true model.  The empirical power of the proposed tests for significance level $\alpha=0.05$ is plotted in Figure~\ref{simMMres}.  Each panel corresponds to a different value of $\sigma_2^2$; for simplicity we report only three values of $\sigma^2_2$ for each value of $N$.  The results for the other values of $\sigma_2^2$ are similar.

The method EqualVC is again the most powerful for this study.  We see similar levels of disparity between EqualVC and Bonferroni as in the simulation study of the previous section.  It is promising that method EqualVC outperforms the method that knows the nuisance random effect.  We can also see that knowing the nuisance random effect has little effect on the power of the test as Bonferroni performs nearly identically to KnownSig1. 

It turns out that it is much easier to detect $\sigma_2$ being non-zero when $\sigma_3$ is small or zero than vice versa; for example, for $N=100$ EqualVC rejects $H_0$ in $93.6\%$ of the simulations when $(\sigma_2^2,\sigma_3^2)=(0.5,0)$, but in only $23.4\%$ of simulations when $(\sigma_2^2,\sigma_3^2)=(0,0.5)$.  Note that the relation between the dimension of $\vecb_2$ and $\vecb_3$ in this setup with $K_x=K_t$ and second-order penalties is $q_3=(q_2/2)^2$.  We also see that method EqualVC does not seem to lose its advantage in power over the other methods when one variance component is non-zero while the other is zero.

The type I error rates varied little as either the sample size or number of basis functions changed.  Averaging over the two samples sizes, for $\alpha=0.05$, the empirical type I error rate was 0.013 for Bonferroni, 0.048 for EqualVC, and 0.019 for KnownSig1.  Given how close its rate is to the nominal level and its strong power performance compared to the other methods in both simulation sections, we recommend using the EqualVC method which assumes a priori $\sigma_2=\sigma_3$ and then conducts a single pseudo-RLRT using the \textcite{greven2008Restricted} approach.  

To further assess the adequacy of the EqualVC method, we also considered using the parametric bootstrap as it is implemented in \texttt{lme4} \autocite{bates2013lme4} for a subset of the $(\sigma_2^2,\sigma_3^2)$ pairs.  We found that using the bootstrap produced similar results to the EqualVC method, with only negligible differences between the two methods in both empirical Type I error and power.  This is similar to the results found in \textcite{scheipl2008Size}.  We take this as further support that the EqualVC approach is sound, and prefer using it to the bootstrap due to its significantly reduced computational overhead.
\section{Analysis of Emissions Data}\label{sec_Emissions}
\begin{figure}
\centering
\includegraphics[width=.75\columnwidth]{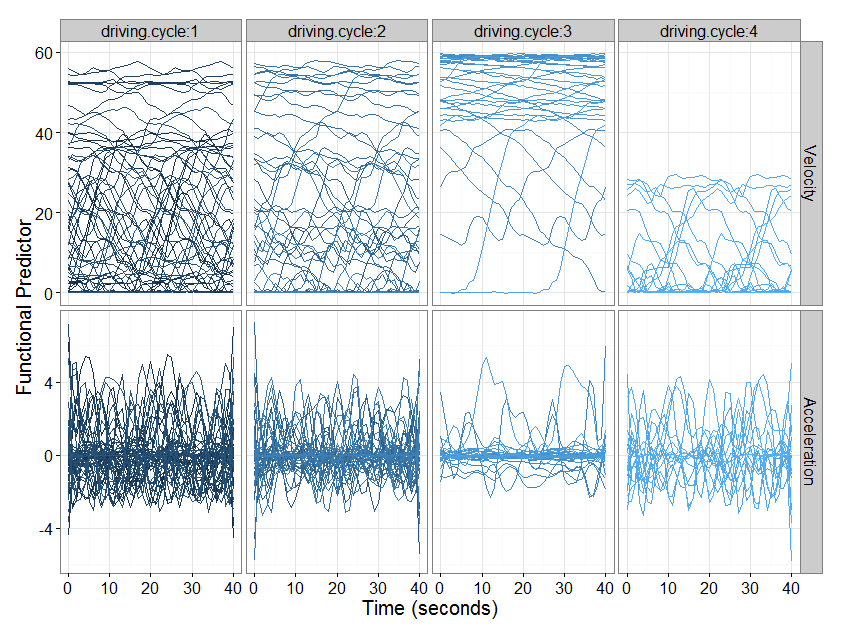}
\caption[Speed and acceleration trajectories for emissions study]{\small Speed and acceleration trajectories in forty seconds blocks grouped according to speed pattern (driving cycle)\label{speed}}
\end{figure}

In this section we apply our proposed procedures to study the quantities of various pollutants in truck exhaust emissions.  The data come from chassis dynamometer emissions readings from the Coordinating Research Council E$55/59$ emissions inventory program \autocite{clark2007emissions}.  The goal of the study was to assess particulate matter emissions in heavy-duty trucks in California.  Vehicles were tested in a lab setting designed to mimic everyday driving conditions.  Particulate matter was captured using 70~mm filters on the dilute exhaust.  Each vehicle in the study was repeatedly driven in four different driving cycles for an extend period.  For example, one driving cycle might be cruising at highway speeds and the next stop-and-go city driving.  For our application, we consider a simplified problem, attempting to model the emissions of one truck from the study.  We attempt to predict the logarithm of particulate matter every twenty seconds using the truck's recorded speed and/or acceleration in the immediately preceding forty seconds.  This resulted in 157 samples.  We chose to sample particulate matter every 20 seconds to ensure there was no temporal dependence between response samples.  Figure~\ref{speed} plots both the original speed data and estimated accelerations for all blocks in the data grouped according to driving cycle.

In the subsection that follows, we analyze the fit of the FLM to these data using our proposed tests for linearity.  After that, we compare FLM and FGAM out-of-sample predictive performance for this data set and also compare predictive performance of the parameterization discussed in Section~\ref{sec_psanova} with the parameterization used in \textcite{McLean2012functional}.
\subsection{FLM Fit Assessment}
\begin{figure}[h]
\centering
\includegraphics[width=.8\columnwidth]{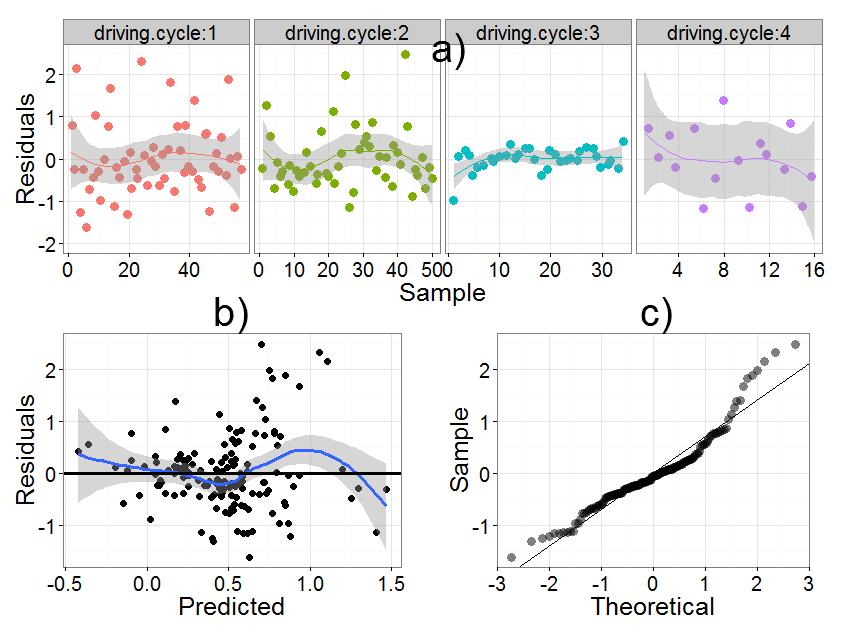}
\caption[Diagnostics for FLM fit to emissions data]{\small Diagnostic plots for an FLM fit with truck acceleration as the functional predictor: a) plots the residuals grouped by driving cycle, b) plots residuals vs. predicted value, and c) is the normal Q-Q plot\label{emissionsFLMres}}
\end{figure}

Leaving the finer details of our fitting procedures to the next section, we now discuss some diagnostics for assessing the fit of both the FLM and the FGAM including use of our proposed testing procedures.  We consider predicting particulate matter using vehicle acceleration as the functional predictor and also include a categorical covariate for the driving cycle.  Some residual plots for an FLM fit to the entire data set using tuning parameters that had been chosen to optimize performance for the next section are given in Figure~\ref{emissionsFLMres}.  The top row of plots shows the residuals grouped according to the driving cycle covariate and the bottom shows the residuals plotted against the predicted value and also a normal Q-Q plot of the residuals.  We can see evidence of a nonlinear association between the residuals and the response and also that the variance of the residuals is not constant across the driving cycle factor.  The Q-Q plot indicates non-normality of the residuals.  
To assess this more formally, we consider the proposed tests of Section~\ref{sec_FGAMtest}.  Using the EqualVC method from Section~\ref{twoVC}, we obtain a p-value $\approx 0$.  This very strongly suggests the FLM is not adequate here.  We also obtain a p-value that is zero to machine precision using the Bonferroni method from Section~\ref{Bonferroni}.  The results remain overwhelming regardless of the number of basis functions used.  

\begin{figure}[h]
\centering
\includegraphics[width=.8\columnwidth]{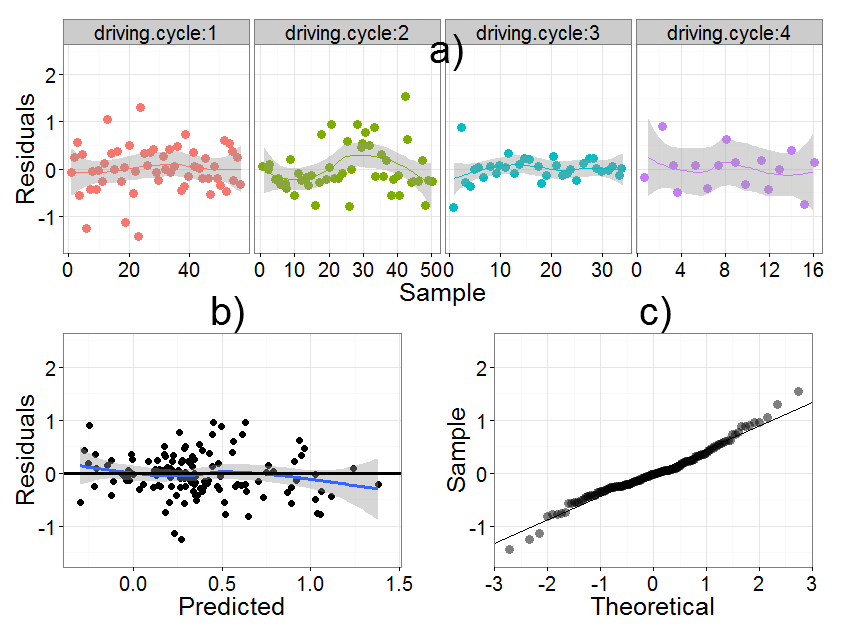}
\caption[Diagnostics for FGAMM fit to emissions data]{\small Diagnostic plots for an FGAMM fit with truck acceleration as the functional predictor: a) plots the residuals grouped by driving cycle, b) plots residuals vs. predicted value, and c) is the normal Q-Q plot\label{emissionsFGAMres}}
\end{figure}

The residual plots for an FGAM fit to the data using the basis construction from this chapter can be seen in Figure~\ref{emissionsFGAMres}.  We can see that the magnitude of the residuals has gone down and that all three plots seem to be less in violation of the model assumptions than the FLM fit.  The variance of the residuals also appears to be more constant across driving cycles.  

\begin{figure}
\centering
\includegraphics[width=.8\columnwidth]{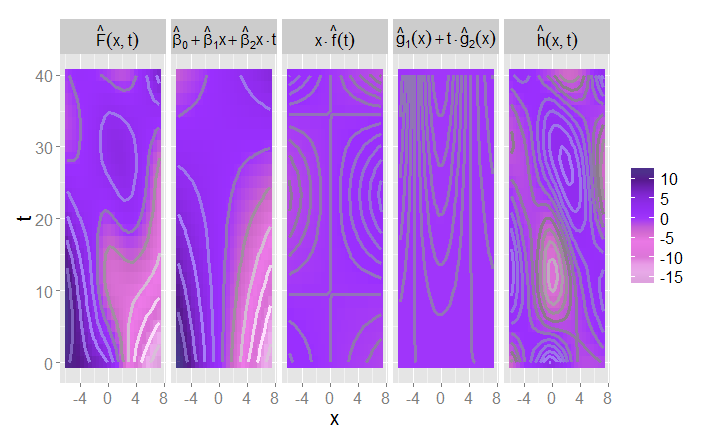}
\caption[Contours of estimated surface for emissions data]{\small Contours of estimated surface, $\hat{F}(x,t)$, and components of FGAMM fit from using acceleration trajectories as predictors.  The second panel is the parametric component of the fit, third is the component parametric in $x$ and nonparametric in $t$, fourth vice versa, and finally $\hat{f}(x,t)$ is nonparametric in both and subject to a fourth order penalty\label{emissionsFit}}
\end{figure}

Figure~\ref{emissionsFit} shows contours of the estimated surface obtained by using all 157 samples and the acceleration curves as predictors.  The surface was estimated using \texttt{lme4} \autocite{bates2013lme4}.  Also plotted are the individual components of the basis construction of Section~\ref{sec_psanova}; the unpenalized component, along with the three penalized components (see Table~\ref{pencomp}).  The marginal bases for the $x$ and $t$ axes were both of dimension eight.  Interestingly, the variance component for the FLM portion of the fit was estimated to be very close to zero in this case.

The FLM fairs only marginally better if the truck speeds are used as the functional predictor.  We omit the diagnostic measures, but the out-of-sample prediction performance using either covariate or both is examined in the next section.
\subsection{Out-of-Sample Prediction of Particulate Matter}
As further confirmation that the FGAM provides a better fit to this data than the FLM, we considered out-of-sample prediction of the log-particulate matter.  We also compare the two different basis constructions for the tensor product surface in our model: the \textcite{McLean2012functional} construction (FGAM) and the construction from the current work (FGAMM).  We also fit the fully nonparametric kernel estimator of \textcite{ferraty2006Nonparametric}.  We considered fitting models with a single functional predictor using either the truck speeds or the accelerations, as well as models including both of these functional predictors at once.  Since multiple functional predictors or scalar covariates do not seem to be implemented for the model in \textcite{ferraty2006Nonparametric}, we could not consider the model with both functional predictors for that method.  It is also for this reason that we do not include a categorical predictor for the driving cycle for any of the models in the reported results.  Perhaps surprisingly, including the categorical predictor for the FLM and FGAM methods had little effect on out-of-sample predictive performance.

\begin{figure}
\centering
\includegraphics[width=0.75\columnwidth]{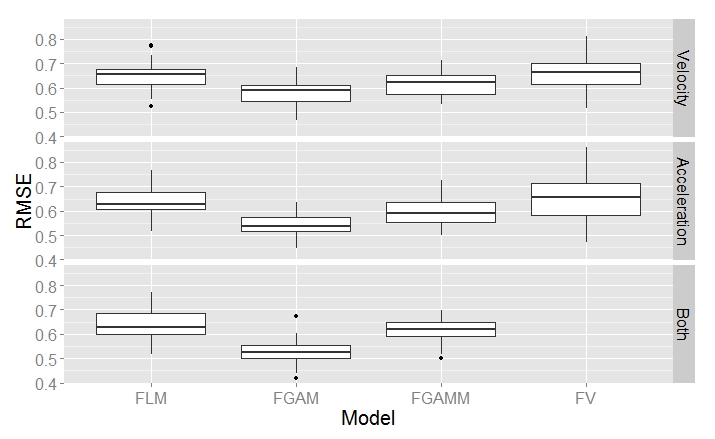}
\caption[Boxplots of prediction error for emissions data]{\small Boxplots for prediction error for FLM, FGAM, FGAMM, and FV using truck speed, acceleration, or both at once as predictors\label{emissionsRMSE}}
\end{figure}

For FLM, FGAM, and FGAMM, smoothing parameters were chosen using REML.  The nonparametric kernel estimator was fit using code from the authors \autocite{staph2006code}, and includes automatic bandwidth selection.  Several different basis dimensions were considered for both the FLM and FGAMs.  Results varied little as the number of basis functions changed for each of the methods, so for compactness we only report the values that produced the lowest root-mean-square error (RMSE) averaged over the different predictors for each method.  For the FLM this was ten basis functions for the functional coefficient, for FGAM this was six basis functions for both axes, and for FGAMM this was ten basis functions for each axis for the one functional predictor models and seven basis functions for the two predictor model.  Both the FGAM and the FLM can be fit in \texttt{R} using the \texttt{refund} package \autocite{refund}.  The underlying estimation is handled by the \texttt{R} recommended package \texttt{mgcv} \autocite{wood2011fast}.  For FGAMM, the variance components are estimated by the package \texttt{lme4} \autocite{bates2013lme4}.  The data was randomly divided so that 105 samples ($\approx$ two thirds of the data) were using for training the models and the remaining samples were used for testing.  Boxplots of the RMSEs for predicting the test set samples over 25 different random partitions into training-test sets is displayed in Figure~\ref{emissionsRMSE}.

We see that both FGAM and FGAMM had lower mean RMSE for the speed, acceleration, and two functional predictor models.  While FGAMM performed similarly to FGAM when truck speed was the functional predictor, the FGAM basis construction provided better out-of-sample predictions for the other two models.  Both parameterizations for the FGAM gave superior prediction results to the Ferraty and Vieu model as well.  The lowest mean RMSE was achieved by the FGAM that included both functional predictors and used the more standard tensor product construction.  A method for continuously predicting emissions over time for the data from this study using a functional response model is considered by \textcite{hooker2012functional}.  A non-functional data approach on an expanded data set from the original study authors can be found in \textcite{clark2010expressing}.
\section{Conclusion}
This work has addressed an important question, which to this point has few answers in the literature: when is a scalar-on-function regression problem not well-modelled by the functional linear model?  How often is the relationship between the response and functional predictor truly nonlinear?  Using an alternative mixed model representation for the recently proposed functional generalized additive model, we were able to develop simple tests for assessing linearity of an FGAM fit to functional data.  Through two simulation studies we were able to find an approach that gave type I error rates quite close to the nominal level and also had high power.  In an application to measuring the amount of pollutants emitted by heavy-duty trucks in various driving conditions, we presented strong evidence that particulate matter could not be adequately predicted from truck speed or acceleration using a functional linear model, whereas the (nonlinear) FGAM provided a much better fit to the data.

Of interest for future work is developing a test where the null model is FGAM to assess if an even more general model, such as the kernel estimator of \textcite{ferraty2006Nonparametric} is necessary.  This is more complicated because the model under the alternative is not a linear mixed model and because there is no explicit nesting of the model under the null hypothesis within the alternative model.  It is also of interest to consider hypothesis testing for models with multiple functional predictors, assessing if adding a second functional covariate significantly improves model fit or perhaps assessing whether there is a significant interaction between two functional predictors.  For example, for the emissions data, including a covariate for speed in a model already containing acceleration seems to offer little improvement in fit; we would like to formally assess this using similar testing procedures to ones developed in this work.  Assessing goodness of fit using Bayes factors for a Bayesian version of FGAM \autocite{mclean2013bayesian} is the subject of ongoing work.
\section*{Acknowledgements}
Much of this work was completed while the first author was a PhD student at Cornell University.  He thanks NSERC for support while completing his studies.  The authors wish to thank Ray Carroll for helpful comments on an early version of the manuscript and Oliver Gao for providing the emissions data.
\printbibliography
\end{document}